\documentclass[letterpaper]{article} 
\usepackage{aaai2026}  
\usepackage{times}  
\usepackage{helvet}  
\usepackage{courier}  
\usepackage[hyphens]{url}  
\usepackage{graphicx} 
\usepackage{natbib}  
\usepackage{caption} 
\usepackage{algorithm}
\usepackage{algorithmic}

\usepackage{newfloat}
\usepackage{listings}
\DeclareCaptionStyle{ruled}{labelfont=normalfont,labelsep=colon,strut=off} 
\floatstyle{ruled}
\newfloat{listing}{tb}{lst}{}
\floatname{listing}{Listing}
\usepackage{amsmath}
\usepackage{amssymb}
\usepackage{xspace}
\usepackage{color}

\usepackage{multirow}
\usepackage{tabularx}
\usepackage{tablefootnote}
\usepackage{booktabs}
\usepackage{makecell}
\usepackage{bbding}

\usepackage{pifont}

\usepackage{microtype}
\usepackage{graphicx}
\usepackage{enumitem}

\usepackage{subfig}

\usepackage{hyperref}

\newcommand{\todo}[1]{\textcolor{black}{#1}}

\newcommand{\tool}{\textit{GenSR-Pref}\xspace}
\newcommand{\anyenhance}{{AnyEnhance}\xspace}
\newcommand{\arse}{{AR+Soundstorm}\xspace}
\newcommand{\flowse}{{Flow-SR}\xspace}

\title{Multi-Metric Preference Alignment for Generative Speech Restoration}
\author {
    Junan Zhang\textsuperscript{\rm 1},
    Xueyao Zhang\textsuperscript{\rm 1},
    Jing Yang\textsuperscript{\rm 2},
    Yuancheng Wang\textsuperscript{\rm 1},
    Fan Fan\textsuperscript{\rm 2},
    Zhizheng Wu\textsuperscript{\rm 1}
}
\affiliations {
    \textsuperscript{\rm 1}Shenzhen Research Institute of Big Data, School of Data Science, The Chinese University of Hong Kong, Shenzhen\\
    \textsuperscript{\rm 2}Central Media Technology Institute, Huawei\\
    \href{mailto:junanzhang@link.cuhk.edu.cn}{junanzhang@link.cuhk.edu.cn}, \href{mailto:wuzhizheng@cuhk.edu.cn}{wuzhizheng@cuhk.edu.cn}
}

\begin{document}

\maketitle

\begin{abstract}
Recent generative models have significantly advanced speech restoration tasks, yet their training objectives often misalign with human perceptual preferences, resulting in suboptimal quality. While post-training alignment has proven effective in other generative domains like text and image generation, its application to generative speech restoration remains largely under-explored. This work investigates the challenges of applying preference-based post-training to this task, focusing on how to define a robust preference signal and curate high-quality data to avoid reward hacking. To address these challenges, we propose a \textbf{multi-metric preference alignment} strategy. We construct a new dataset, \textbf{\tool}, comprising \todo{80K} preference pairs, where each chosen sample is unanimously favored by a complementary suite of metrics covering perceptual quality, signal fidelity, content consistency, and timbre preservation. This principled approach ensures a holistic preference signal. Applying Direct Preference Optimization (DPO) with our dataset, we observe consistent and significant performance gains across three diverse generative paradigms: autoregressive models (AR), masked generative models (MGM), and flow-matching models (FM) on various restoration benchmarks, in both objective and subjective evaluations. Ablation studies confirm the superiority of our multi-metric strategy over single-metric approaches in mitigating reward hacking. Furthermore, we demonstrate that our aligned models can serve as powerful ``data annotators'', generating high-quality pseudo-labels to serve as a supervision signal for traditional discriminative models in data-scarce scenarios like singing voice restoration.
\end{abstract}

\begin{links}
    \link{Demopage}{https://gensr-pref.github.io}
\end{links}


\section{Introduction}
\label{sec:intro}

Speech restoration, which aims to recover high-quality speech from various degradations, is a fundamental task in audio processing~\cite{liu2022voicefixer,zhang24h_interspeech}. Recent advances in utilizing generative models for speech restoration, namely \textit{generative speech restoration (GenSR)}, have shown remarkable performance in tasks such as denoising, dereverberation, declipping, super-resolution, etc.~\cite{zhang2025anyenhance,yang2024genhancer,li2024masksr,wang2025metis,wang2024selm,anonymous2025gense,kang2025llase,wang2024speechx,liu2024audiosr}. Unlike traditional discriminative methods that minimize a distance metric to a clean reference, generative speech restoration models are trained to maximize the likelihood of clean speech given a degraded input, thus learning the underlying distribution of clean speech~\cite{lemercier2023analysing}, enabling them to generate high-fidelity audio even from severely degraded inputs.

In parallel, post-training techniques, designed to align generative models with specific downstream objectives or human preferences, have become integral to the advancement of generative modeling in some domains. In natural language processing~\cite{ouyang2022instructgpt,rafailov2023direct,bai2022rlhf}, text-to-image/video~\cite{xu2023imagereward,wallace2024diffusiondpo,liu2025videoalign}, and text-to-speech/audio~\cite{zhang2025advancing,sun2025f5r,hussain2025koel,tian2025preference,yao2025fine,zhang2024speechalign,liao2024baton} synthesis, preference-based alignment methods have been instrumental in enhancing model quality, safety, and human alignment. However, despite its potential, the application of such post-training alignment to the domain of generative speech restoration remains largely under-explored.

Applying preference-based post-training to GenSR models presents a unique set of challenges. Chief among them are: (1) \textbf{Defining a faithful preference signal}: How to construct an accessible, automated proxy that captures the multi-faceted nature of human auditory perception (which values clarity, naturalness, and a lack of artifacts)? (2) \textbf{Curating high-quality preference data}: Given a preference signal, what is an effective strategy for constructing preference pairs that can robustly guide model optimization? (3) \textbf{Mitigating reward hacking}: How to ensure the model achieves holistic, genuine improvement, rather than simply learning to exploit the biases of a specific metric?

To address these challenges, we propose and investigate a \textbf{multi-metric preference alignment} strategy. We argue that a robust solution to reward hacking lies in the preference signal itself being multi-dimensional and holistic. To this end, we construct a new preference dataset, which we name \textbf{\tool}. A sample is chosen as winner only when it is \textit{unanimously} judged superior by a complementary suite of metrics that assess four distinct aspects of quality: perceptual quality, signal-level fidelity, content alignment, and timbre preservation. This strict, unanimous criterion ensures that our model learns from a robust and holistic preference signal, effectively mitigating the risk of reward hacking.

Based on the \tool\ dataset, we apply Direct Preference Optimization (DPO)~\cite{rafailov2023direct} to align models across three major distinct generative paradigms: the sequential \textbf{Autoregressive (AR)} models, the iterative \textbf{Masked Generative (MGM)} models (also known as masked or discrete diffusion), and the continuous \textbf{Flow-Matching (FM)} models. Our approach yields significant and consistent improvements across all benchmarks \todo{even with only 3k pairs}, confirmed by objective and subjective tests.
Our key findings are threefold. First, our multi-metric signal is crucial for avoiding the \textbf{reward hacking} inherent in single-metric optimization. Second, we show that using ground-truth as a fixed winner causes \textbf{model collapse}, highlighting the need for learned, relative preferences over absolute targets. Third, models perform best with data from their own architecture, a principle we term \textbf{in-paradigm alignment}. We attribute this to unique ``alignment directions'' for each paradigm, where in-paradigm data provides a more direct optimization path. Finally, we demonstrate that our aligned models can act as powerful ``data annotators'', generating high-quality pseudo-labels to train discriminative models in data-scarce scenarios like singing voice restoration, thus bridging the gap between generative and discriminative paradigms.

To summarize, our main contributions are:
\begin{itemize}
    \item We propose a \textbf{multi-metric preference alignment} strategy to address the challenges of defining a robust preference signal and mitigating reward hacking in post-training for generative speech restoration. Through extensive experiments, we demonstrate that this strategy yields significant and consistent improvements across three diverse generative paradigms (AR, MGM, and FM).
    \item We construct and introduce \textbf{\tool}, a dataset of \todo{80k} preference pairs built upon a strict, \textbf{unanimous multi-metric agreement criterion}. This dataset, which captures a holistic aspect of audio quality, will be publicly released to facilitate future research.
    \item We showcase a practical application of our aligned models, demonstrating their ability to serve as ``data annotators'' to generate pseudo-labels, thereby empowering the training of traditional discriminative models in data-scarce scenarios.
\end{itemize}

\section{Related Work}

\subsection{Generative Speech Restoration}

The paradigm for speech restoration has recently shifted towards generative modeling, which can be broadly categorized by the underlying generation mechanism. One major stream leverages \textbf{autoregressive language models} to treat restoration as a sequence-to-sequence problem, as seen in SELM~\cite{wang2024selm}, TSELM~\cite{tang2024tselm}, GenSE~\cite{anonymous2025gense}, and versatile multi-task systems like SpeechX~\cite{wang2024speechx}, UniAudio~\cite{Yang2024UniAudio}, and LLaSE-G1~\cite{kang2025llase}. A second family consists of \textbf{masked generative models}, or discrete diffusion models, which iteratively refine tokenized audio representations. This approach has been applied to general restoration (e.g., MaskSR~\cite{li2024masksr, liu2024joint}, AnyEnhance~\cite{zhang2025anyenhance}), speech enhancement (e.g., Genhancer~\cite{yang2024genhancer}), and target speaker extraction (e.g., Metis~\cite{wang2025metis}). A third direction employs \textbf{generative models with continuous dynamics} to learn mappings in a continuous space; this includes methods based on score-SDE~\cite{welker2022sgmse}, latent diffusion~\cite{liu2024audiosr, wang2025solospeech}, flow-matching~\cite{wang2025flowse, liugenerative, ku2024generative}, and schr{\"o}dinger bridges~\cite{li2025bridgesr, jukic2024bridgese} for various restoration tasks. Despite achieving state-of-the-art results, these generative models are typically trained using likelihood-based objectives, which often misalign with human perceptual quality. Our work addresses this gap by introducing a preference-based post-training strategy applicable across these diverse generative paradigms.

\subsection{Post-Training for Audio Alignment}

Post-training alignment has proven effective across various generative domains, including text~\cite{ouyang2022instructgpt,rafailov2023direct,bai2022rlhf}, vision~\cite{xu2023imagereward,wallace2024diffusiondpo,liu2025videoalign}, speech~\cite{zhang2025advancing,sun2025f5r,hussain2025koel,tian2025preference,yao2025fine,zhang2024speechalign}, music~\cite{cideron2024musicrl,lei2025levo}, and audio effects~\cite{majumder2024tango2,liao2024baton}. In the context of speech restoration, early efforts like MetricGAN~\cite{fu2019metricgan} used adversarial training to optimize for metrics like PESQ. More recently, preference-based methods have emerged, such as aligning with a learned NISQA predictor via PPO~\cite{kumar2025using} or using DPO with a single metric like UTMOS~\cite{li2025aligning}. While single-metric alignment is effective, it risks reward hacking by ignoring that robust restoration is multi-faceted, requiring a balance of quality, fidelity, and timbre (Section~\ref{sec:ablation}). Grounded in this principle, we propose a \textbf{multi-metric alignment strategy} to pursue holistic improvement. We systematically show this approach yields more comprehensive gains across diverse generative paradigms while naturally mitigating reward hacking.

\section{Multi-Meric Preference Alignment}
\label{sec:method}

\begin{figure}[htbp]
    \centering
    \includegraphics[width=0.98\linewidth]{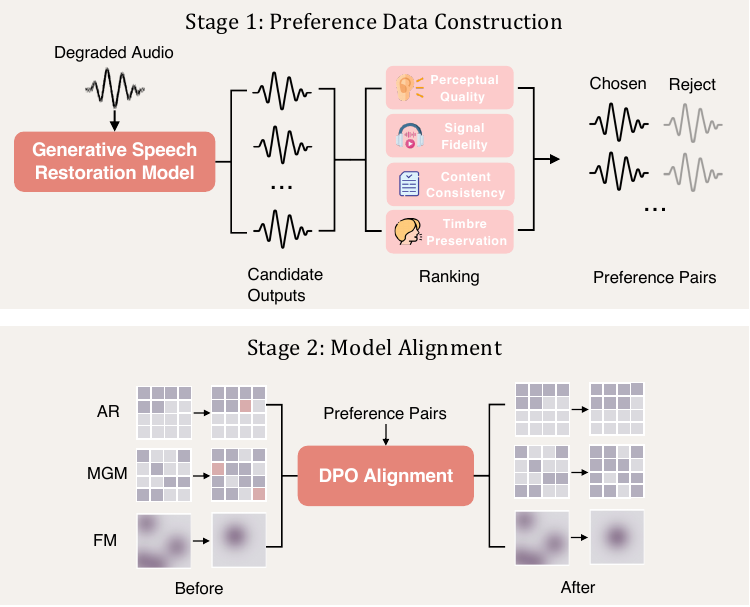}
    \caption{An overview of our multi-metric preference alignment strategy. The process consists of two main stages: (1) constructing the \tool\ preference dataset by ranking model outputs based on a unanimous agreement across multiple metrics, and (2) fine-tuning the model with these preferences using Direct Preference Optimization (DPO).}
    \label{fig:overview}
\end{figure}

Our approach to aligning generative speech restoration models is centered around a \textbf{multi-metric preference alignment} strategy, which is illustrated in Figure~\ref{fig:overview}. This strategy involves two key stages: first, the construction of a high-quality preference dataset, and second, the application of Direct Preference Optimization (DPO) to align the model with these preferences. This approach is designed to be model-agnostic, allowing it to be applied to any generative restoration model.

\subsection{The \tool\ Dataset: Curating a Holistic Preference Signal}
\label{subsec:data_construction}

A pivotal element of our alignment strategy is the \textbf{\tool} dataset, designed to provide a robust, holistic preference signal while mitigating single-metric reward hacking. To ensure our dataset's breadth and quality, we generate candidate outputs using models from three distinct generative paradigms. We then form preference pairs by ranking outputs exclusively within each paradigm. This approach is chosen to capture the subtle variations in a single model's generations, providing a fine-grained signal tailored to its specific error patterns. This comprehensive strategy yields a dataset for robust alignment and enables detailed in-paradigm versus cross-paradigm analyses.

\subsubsection{Model Selection}

To comprehensively evaluate our alignment framework across different generative paradigms, we adopt one representative model from each of the three major families: autoregressive (AR), masked generative (MGM), and flow-matching (FM) models.

\noindent\textbf{Masked Generative Model (MGM).}  
We use the pre-trained AnyEnhance~\cite{zhang2025anyenhance} model as our MGM model. It predicts acoustic tokens based on the DAC codec~\cite{kumar2024high} from partially masked sequences, effectively capturing non-autoregressive and context-aware dependencies.

\noindent\textbf{Autoregressive Model (AR).}  
We train a new model, \arse, which follows a two-stage pipeline common in text-to-speech (TTS) domain. It first predicts semantic tokens of the Metis codec~\cite{wang2025metis} given noisy inputs, and a second-stage non-autoregressive model similar to Soundstorm~\cite{borsos2023soundstorm} then converts these tokens into acoustic tokens for waveform synthesis.

\noindent\textbf{Flow-Matching Model (FM).}  
We train \flowse, a model that learns to predict the velocity field of clean mel-spectrograms from noisy inputs via optimal transport interpolation based on the Diffusion Transformer (DiT) architecture~\cite{peebles2023scalable}. A pre-trained Vocos vocoder~\cite{siuzdak2023vocos} is then used to synthesize the final waveform from the generated spectrograms. \todo{Details on model architectures and training procedures can be found in Appendix~\ref{app:implementation_details}.}

To ensure a reliable preference signal that captures the multi-faceted nature of audio quality, we evaluate each candidate output across four complementary dimensions:

\begin{itemize}
    \item \textbf{Perceptual Quality (NISQA)}~\cite{mittag2021nisqa}: Assesses overall listening quality, including naturalness and the absence of annoying artifacts.
    \item \textbf{Signal-level Fidelity (DNSMOS)}~\cite{reddy2022dnsmos}: A composite metric evaluating signal distortion, background noise, and overall quality for a fine-grained assessment of signal integrity.
    \item \textbf{Content Alignment (SpeechBERTScore)}~\cite{saeki2024speechbertscore}: Measures semantic similarity to the ground-truth transcription to ensure linguistic content is not altered during restoration.
    \item \textbf{Timbre Preservation (Speaker Similarity)}: Uses a pre-trained speaker verification model to compute cosine similarity, ensuring the speaker's identity is preserved.
\end{itemize}

A preference pair $(y_w, y_l)$ is formed only under a \textbf{strict, unanimous agreement criterion}: the winning output $y_w$ must score higher than the losing output $y_l$ on \textit{all} of these metrics simultaneously. This process ensures each preference pair represents a holistic and unambiguous improvement, providing a robust signal for the DPO alignment stage.

In total, \tool comprises approximately 80K preference pairs, structured for distinct experimental goals. The primary component is a large-scale subset of \textbf{69,456} pairs from the \anyenhance (MGM) model, allowing us to validate our alignment strategy's effectiveness with a substantial amount of data. For fair comparisons and ablation studies—particularly our cross-paradigm analysis—we created smaller, controlled subsets from a shared set of input utterances, yielding \textbf{3,354} pairs for AR, \textbf{3,428} for FM, and \textbf{3,035} for MGM. This dual-component structure enables both an investigation into data scaling effects and rigorous, controlled analyses of alignment dynamics.

\subsection{GenSR Model Alignment via Direct Preference Optimization}
\label{subsec:dpo_alignment}

In the second stage of our approach, we align the generative speech restoration models using the \tool\ dataset. We employ \textbf{Direct Preference Optimization (DPO)}~\cite{rafailov2023direct}, a simple yet powerful technique that directly optimizes a policy to satisfy preferences without explicit reward modeling or reinforcement learning. The core idea of DPO is to re-parameterize the reward function in the standard Reinforcement Learning from Human Feedback (RLHF) objective using the optimal policy, which leads to a simple contrastive loss over preferred and dispreferred samples. In this section, we detail how we adapt and apply DPO to the three diverse generative paradigms used in our study: Autoregressive (AR), Masked Generative (MGM), and Flow-Matching (FM) models. \todo{Detailed derivations can be found in Appendix~\ref{app:dpo_derivation}.}

\subsubsection{DPO for Autoregressive Models}

AR models treat speech restoration as a sequence-to-sequence task. Given a degraded input $x$, they generate a clean audio sequence $y = (y_1, ..., y_T)$ token by token. They are typically trained by maximizing the log-likelihood of the ground-truth sequence, using a standard cross-entropy loss:
\begin{equation}
    \scriptsize
    \mathcal{L}_{\text{AR}} = - \mathbb{E}_{(x, y) \sim \mathcal{D}_{\text{train}}} \left[ \sum_{t=1}^{T} \log p_\theta(y_t | y_{<t}, x) \right]
\end{equation}

\noindent\textbf{DPO Derivation.} The standard RLHF objective aims to find a policy $p_\theta$ that maximizes the reward from a learned reward model $r_\phi(x, y)$, while constraining its deviation from a reference policy $p_\text{ref}$ using a KL-divergence term:
\begin{equation}
    \scriptsize
    \max_{p_\theta} \, \mathbb{E}_{y \sim p_\theta(y|x)}[r_\phi(x, y)] - \beta D_{\text{KL}}\left[ p_\theta(y|x) \,\|\, p_{\text{ref}}(y|x) \right]
\end{equation}
The key insight of DPO is that for this objective, the optimal policy $p^*_\theta$ has a closed-form solution: $p^*_\theta(y|x) \propto p_{\text{ref}}(y|x) \exp(\frac{1}{\beta}r_\phi(x,y))$. By rearranging this equation, the unknown reward function $r_\phi(x,y)$ can be expressed in terms of the optimal policy and the reference policy. Substituting this back into the Bradley-Terry loss for the reward model, which models the probability of preferring $y_w$ over $y_l$, yields the final DPO loss that directly optimizes the policy $p_\theta$:
\begin{equation}
    \scriptsize
    \label{eq:dpo_ar}
    \begin{aligned}
        \mathcal{L}_{\text{DPO}} = - \mathbb{E}_{(x, y_w, y_l) \sim \mathcal{D}} \Bigg[ \log \sigma \Big(
            \beta \log \frac{p_\theta(y_w|x)}{p_{\text{ref}}(y_w|x)} - \beta \log \frac{p_\theta(y_l|x)}{p_{\text{ref}}(y_l|x)} \Big) \Bigg]
    \end{aligned}
\end{equation}
where $\mathcal{D}$ is our \tool\ preference dataset, and $\sigma$ is the sigmoid function.

\subsubsection{DPO for Masked Generative Models}
\label{sssec:dpo_mgm}

MGM models~\cite{chang2022maskgit,zhang2025anyenhance,wang2025metis,ju2024naturalspeech,wang2024maskgct}, also known as discrete diffusion models, learn to restore a clean speech sequence $y_0$ from a partially masked or corrupted version $y_t$, where $t \in [0, T]$ is the noise step parameter that controls the amount of corruption, and $y_T$ is the fully masked sequence. They are trained to predict the original tokens in the masked positions, optimizing a masked language modeling objective:
\begin{equation}
    \scriptsize
    \mathcal{L}_{\text{MGM}} = - \mathbb{E}_{(x, y_0), t, m_t} \left[ \sum_{i=1}^{T} m_{t,i} \log p_\theta(y_{0,i} | y_t, x) \right]
\end{equation}
where $m_t \in \{0, 1\}^T$ is a binary mask indicating which tokens in the sequence are masked (1) or unmasked (0). The model learns to predict the original tokens in the masked positions, effectively learning to denoise the corrupted input.
\noindent\textbf{DPO Derivation.} Following the INTP framework~\cite{zhang2025advancing}, we extend DPO to this non-autoregressive setting. The core logic remains the same, but the policy now represents the conditional probability of the full clean sequence $y_0$ given the corrupted version $y_t$. The DPO loss for MGM is thus analogous to the AR case, contrasting the log-probabilities of the preferred and dispreferred sequences, conditioned on their respective masked inputs:

\begin{equation}
    \scriptsize
    \label{eq:dpo_mgm}
    \begin{aligned}
        \mathcal{L}_{\text{DPO-MGM}} = - \mathbb{E}_{(x, y_w, y_l) \sim \mathcal{D},\, t} \Bigg[ \log \sigma \Big(
        \beta \log \frac{p_\theta(y_0^w|y_t^w, x)}{p_{\text{ref}}(y_0^w|y_t^w, x)} \\
        {} - \beta \log \frac{p_\theta(y_0^l|y_t^l, x)}{p_{\text{ref}}(y_0^l|y_t^l, x)} \Big) \Bigg]
    \end{aligned}
\end{equation}
where $y_t^w$ and $y_t^l$ are the masked versions of the winning ($y_0^w$) and losing ($y_0^l$) sequences from our preference dataset.

\subsubsection{DPO for Flow-Matching Models}
\label{sssec:dpo_fm}

FM models~\cite{lipman2022flow} learn a continuous mapping from a simple prior distribution (e.g., Gaussian noise $y_0 \sim \mathcal{N}(0, I)$) to a complex data distribution (e.g., clean speech $y_1$). They are trained to predict the velocity vector field $v_\theta(y_t, t, x)$ that describes the probabilistic flow of data points along a linear interpolation path from the noise $y_0$ to the data $y_1$. The standard training objective minimizes the difference between the predicted velocity and the true velocity:
\begin{equation}
    \scriptsize
    \mathcal{L}_{\text{FM}} = \mathbb{E}_{t, y_0, y_1, x} \left[ \| v_\theta(y_t, t, x) - (y_1 - y_0) \|_2^2 \right]
\end{equation}
where $y_t = (1-t)y_0 + ty_1$ is a point on the linear trajectory from the noise to the data, and $t$ is a noise step parameter that controls the interpolation between the two. The model learns to predict the velocity field that describes the continuous flow of data points along the linear interpolation path, effectively learning to denoise by predicting the gradient of the data distribution along the linear interpolation.

\noindent\textbf{DPO Derivation.} To extend DPO to FM models, we follow recent advances in aligning diffusion-based generators~\cite{wallace2024diffusiondpo,liu2025videoalign}. Instead of modeling the full data likelihood along the entire generative trajectory, we use a tractable surrogate objective that operates on a single noise step. Specifically, we approximate the DPO objective using the model's instantaneous prediction error at a sampled timestep, treating the L2 loss between predicted and true velocities as a proxy for negative log-likelihood. The resulting loss encourages the model to reduce velocity error for preferred samples ($y_w$) while increasing it for dispreferred ones ($y_l$), yielding the following expression:
\begin{equation}
    \scriptsize
    \label{eq:dpo_fm}
    \mathcal{L}_{\text{DPO-FM}} = - \mathbb{E}_{(x, y_w, y_l) \sim \mathcal{D}, t, y_0} \left[ \log \sigma\left(-\beta (\Delta_w - \Delta_l)\right) \right]
\end{equation}
where $\Delta$ represents the difference in squared L2 error between the trained model $v_\theta$ and the reference model $v_{\text{ref}}$ for a given sample ($y_w$ or $y_l$):
\begin{equation}
    \scriptsize
    \Delta_w = \|v_\theta(y_t^w, t, x) - (y_1^w - y_0)\|_2^2 - \|v_{\text{ref}}(y_t^w, t, x) - (y_1^w - y_0)\|_2^2
\end{equation}
and $\Delta_l$ is defined analogously for the losing sample $y_l$. This single-step formulation provides a scalable and effective approximation to DPO in diffusion processes.

\section{Experiments}

\subsection{Experimental Setup}

\noindent\textbf{Models and Datasets.}
We evaluate our multi-metric preference alignment strategy across the three generative paradigms: \anyenhance for MGM, \arse for AR, and \flowse for FM. All alignment experiments are conducted using our newly constructed \tool dataset. Specifically, we conduct the main experiment (Section~\ref{sec:result-main}) using the primary preference subset (69k pairs for MGM, approx. 3k for AR/FM), while the ablations (Section~\ref{sec:ablation}) are conducted using the controlled smaller 3k-pair subsets for all models to ensure fair and direct comparisons.

\begin{table*}[htbp]
    \centering
    \small
    \renewcommand{\arraystretch}{1}
    \begin{tabular}{cllcccccccc}
        \toprule
        \textbf{Dataset} & \textbf{Model} & \textbf{Type} & \makecell{\textbf{DPO-}\\\textbf{aligned?}} & \textbf{SIG$\uparrow$} & \textbf{BAK$\uparrow$} & \textbf{OVRL$\uparrow$} & \textbf{NISQA$\uparrow$} & \makecell{\textbf{Speech-}\\\textbf{BERTScore$\uparrow$}} & \textbf{Similarity$\uparrow$} \\
        \midrule
        
        \multirow{11}{*}{Voicefixer-GSR}
        & Voicefixer & DISC. & - & 3.299 & 3.971 & 3.003 & 4.160 & 0.797 & 0.882 \\
        & MaskSR & MGM & - & 3.445 & 3.971 & 3.128 & - & - & - \\
        \cmidrule{2-10}
        & \multirow{2}{*}{\anyenhance} & \multirow{2}{*}{MGM} & \ding{55} & 3.406 & 4.073 & 3.136 & 4.308 & 0.829 & 0.924 \\
        &&& \ding{51} & \textbf{3.532} & \textbf{4.091} & \textbf{3.267} & \textbf{4.639} & \textbf{0.834} & \textbf{0.935} \\
        \cmidrule{2-10}
        & \multirow{2}{*}{\arse} & \multirow{2}{*}{AR} & \ding{55} & 3.550 & 4.097 & 3.294 & 4.556 & 0.788 & 0.894 \\
        &&& \ding{51} & \textbf{3.564} & \textbf{4.144} & \textbf{3.331} & \textbf{4.850} & \textbf{0.803} & \textbf{0.904} \\
        \cmidrule{2-10}
        & \multirow{2}{*}{\flowse} & \multirow{2}{*}{FM} & \ding{55} & 3.398 & 3.969 & 3.104 & 4.010 & 0.812 & 0.918 \\
        &&& \ding{51} & \textbf{3.483} & \textbf{4.092} & \textbf{3.230} & \textbf{4.672} & \textbf{0.830} & \textbf{0.924} \\
        
        \midrule
        \multirow{8}{*}{Librivox-GSR}
        & \multirow{2}{*}{\anyenhance} & \multirow{2}{*}{MGM} & \ding{55} & 3.546 & 4.142 & 3.308 & 4.346 & 0.822 & \textbf{0.955} \\
        &&& \ding{51} & \textbf{3.690} & \textbf{4.201} & \textbf{3.475} & \textbf{4.865} & \textbf{0.828} & 0.954 \\
        \cmidrule{2-10}
        & \multirow{2}{*}{\arse} & \multirow{2}{*}{AR} & \ding{55} & 3.663 & 4.134 & 3.419 & 4.535 & 0.783 & 0.922 \\
        &&& \ding{51} & \textbf{3.693} & \textbf{4.193} & \textbf{3.478} & \textbf{4.923} & \textbf{0.793} & \textbf{0.924} \\
        \cmidrule{2-10}
        & \multirow{2}{*}{\flowse} & \multirow{2}{*}{FM} & \ding{55} & 3.550 & 4.062 & 3.281 & 4.184 & 0.791 & \textbf{0.931} \\
        &&& \ding{51} & \textbf{3.602} & \textbf{4.152} & \textbf{3.368} & \textbf{4.825} & \textbf{0.801} & 0.930 \\
        
        \midrule
        \multirow{8}{*}{CCMusic-GSR}
        & \multirow{2}{*}{\anyenhance} & \multirow{2}{*}{MGM} & \ding{55} & 3.243 & 3.547 & 2.797 & 3.345 & 0.811 & \textbf{0.915} \\
        &&& \ding{51} & \textbf{3.440} & \textbf{3.827} & \textbf{3.062} & \textbf{4.154} & \textbf{0.817} & 0.909 \\
        \cmidrule{2-10}
        & \multirow{2}{*}{\arse} & \multirow{2}{*}{AR} & \ding{55} & 3.378 & 3.693 & 2.956 & 3.948 & 0.710 & \textbf{0.854} \\
        &&& \ding{51} & \textbf{3.460} & \textbf{3.865} & \textbf{3.094} & \textbf{4.438} & \textbf{0.713} & 0.853 \\
        \cmidrule{2-10}
        & \multirow{2}{*}{\flowse} & \multirow{2}{*}{FM} & \ding{55} & 3.298 & 3.540 & 2.813 & 3.897 & 0.733 & \textbf{0.883} \\
        &&& \ding{51} & \textbf{3.378} & \textbf{3.770} & \textbf{2.971} & \textbf{4.371} & \textbf{0.740} & 0.881 \\

        \bottomrule
    \end{tabular}
    \caption{Quantitative comparison on GSR benchmarks across three generative paradigms—MGM (\anyenhance), AR (\arse), and FM (\flowse)—before and after DPO alignment. Results from a discriminative baseline Voicefixer (DISC.) and the MaskSR model are also included. Bold indicates improvements after alignment.}
    \label{tab:main-results-gsr}
\end{table*}

\noindent\textbf{Evaluation Benchmarks and Metrics.}
We conduct evaluations on two major kinds of benchmarks: (1) General speech restoration (GSR) that includes denoising, dereveration, declipping and super-resolution at the same time (Voicefixer-GSR, Librivox-GSR, CCMusic-GSR), (2) Speech Enhancement (SE) under noisy and reverberant conditions (DNS-No-Reverb and DNS-With-Reverb). For a comprehensive evaluation, we not only include the original unaligned models, but also involve state-of-the-art baselines LLaSE-G1~\cite{kang2025llase}, GenSE~\cite{anonymous2025gense}, MaskSR~\cite{li2024masksr}, and FlowSE~\cite{wang2025flowse}, and a discriminative model Voicefixer~\cite{liu2022voicefixer}. Performance is measured using a comprehensive suite of objective metrics: DNSMOS (SIG, BAK, OVRL)~\cite{reddy2022dnsmos} for signal fidelity and quality, NISQA~\cite{mittag2021nisqa} for perceptual quality, SpeechBERTScore (SBERT)~\cite{saeki2024speechbertscore} for content consistency, and speaker similarity (SIM). For subjective evaluation, we conduct A/B preference tests to assess human perception. \todo{Details on the evaluation can be found in Appendix~\ref{app:evaluation_details}.}

\subsection{Main Results: Effectiveness of Multi-Metric Preference Alignment}
\label{sec:result-main}

\begin{figure}[htbp]
    \centering
    \includegraphics[width=0.9\linewidth]{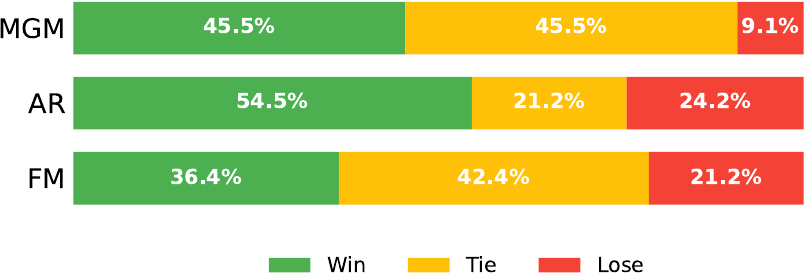}
    \caption{
        Human AB preference test results for DPO-aligned models on Librivox-GSR testset.
    }
    \label{fig:abtest}
\end{figure}

Table~\ref{tab:main-results-gsr} presents the main results, showing that our multi-metric preference alignment strategy is consistently effective across all three paradigms. Notably, the MGM model achieves significant gains when aligned on its large-scale 69k-pair subset (e.g., \textbf{+0.519} NISQA on LibriVox-GSR), demonstrating our method's scalability. Impressively, the AR and FM models also show significant improvements (e.g., \textbf{+0.388} and \textbf{+0.641} NISQA) even with their much smaller, 3k-pair data, highlighting the data efficiency of our approach. These objective gains are confirmed by subjective A/B tests (Figure~\ref{fig:abtest}), where human listeners consistently preferred the aligned models, with win rates reaching up to \textbf{54.5\%}.

Moreover, this strong performance generalizes beyond GSR tasks. As shown in Table~\ref{tab:result-se}, models aligned with our diverse preference data also exhibit seamless performance gains on the downstream DNS speech enhancement benchmark. For example, the aligned \anyenhance model boosts its OVRL score from 3.204 to 3.438 on the reverberant set, demonstrating the robustness of our alignment strategy. \todo{More results and audio visualizations can be found in Appendix~\ref{app:more_results}.}

\begin{table}[htbp]
    \centering
    \small
    \setlength{\tabcolsep}{0.8mm}
    \renewcommand{\arraystretch}{1}
    \begin{tabular}{cllcccc}
        \toprule
        \textbf{Data} & \textbf{Model} & \textbf{Type} & \textbf{Aligned?} & \textbf{SIG} & \textbf{BAK} & \textbf{OVRL} \\
        \midrule

        \multirow{10}{*}{\makecell{No\\ Reverb}}
        & MaskSR & MGM & - & 3.586 & 4.116 & 3.339 \\
        & GenSE & AR & - & 3.650 & 4.180 & 3.430 \\
        & LLaSE-G1 & AR & - & 3.660 & 4.170 & 3.420 \\
        & FlowSE & FM & - & 3.685 & 4.201 & 3.445 \\
        \cmidrule{2-7}
        & \multirow{2}{*}{\anyenhance} & \multirow{2}{*}{MGM} & \ding{55} & 3.640 & 4.179 & 3.418 \\
        &&& \ding{51} & \textbf{3.684} & \textbf{4.203} & \textbf{3.476} \\
        & \multirow{2}{*}{\arse} & \multirow{2}{*}{AR} & \ding{55} & 3.648 & 4.155 & 3.422 \\
        &&& \ding{51} & \textbf{3.673} & \textbf{4.194} & \textbf{3.469} \\
        & \multirow{2}{*}{\flowse} & \multirow{2}{*}{FM} & \ding{55} & 3.581 & 4.133 & 3.355 \\
        &&& \ding{51} & \textbf{3.632} & \textbf{4.173} & \textbf{3.420} \\
        
        \midrule
        
        \multirow{10}{*}{\makecell{With\\ Reverb}}
        & MaskSR & MGM & - & 3.531 & 4.065 & 3.253 \\
        & GenSE & AR & - & 3.490 & 3.730 & 3.190 \\
        & LLaSE-G1 & AR & - & 3.590 & 4.100 & 3.330 \\
        & FlowSE & FM & - & 3.601 & 4.102 & 3.331 \\
        \cmidrule{2-7}
        & \multirow{2}{*}{\anyenhance} & \multirow{2}{*}{MGM} & \ding{55} & 3.500 & 4.040 & 3.204 \\
        &&& \ding{51} & \textbf{3.670} & \textbf{4.178} & \textbf{3.438} \\
        & \multirow{2}{*}{\arse} & \multirow{2}{*}{AR} & \ding{55} & 3.681 & 4.127 & 3.431 \\
        &&& \ding{51} & \textbf{3.709} & \textbf{4.189} & \textbf{3.496} \\
        & \multirow{2}{*}{\flowse} & \multirow{2}{*}{FM} & \ding{55} & 3.539 & 4.019 & 3.255 \\
        &&& \ding{51} & \textbf{3.629} & \textbf{4.163} & \textbf{3.399} \\

        \bottomrule
    \end{tabular}
    \caption{Evaluation results on the DNS speech enhancement benchmark. Models aligned on diverse restoration preference pairs (e.g., denoising, declipping) exhibit seamless performance gains when applied to speech enhancement as a downstream task.}
    \label{tab:result-se}
\end{table}

\subsection{Ablation Studies and Analysis}
\label{sec:ablation}

To better understand the factors contributing to our performance gains, we conduct a series of in-depth ablation studies.

\noindent\textbf{Multi-Metric vs. Single-Metric Alignment.}
To validate our core hypothesis that a multi-metric signal is crucial for avoiding reward hacking, we compare our Multi-Metric Preference Alignment approach with DPO alignment using preference pairs constructed from single metrics. As shown in Table~\ref{tab:metric_ablation}, while optimizing for a single metric improves that specific score, it often leads to stagnation or even degradation in other, un-targeted metrics. For example, the Similarity-aligned model performs worse than the baseline on OVRL and SIG. In contrast, our multi-metric approach achieves robust improvements across all metrics simultaneously. This confirms that our unanimous agreement criterion is effective at mitigating reward hacking and promoting holistic quality improvement. \todo{More results can be found in Appendix~\ref{app:multi_metric_ablation}.}

\begin{table}[h]
    \centering
    \small
    \setlength{\tabcolsep}{1.2mm}
    \renewcommand{\arraystretch}{1}
    \begin{tabular}{lccccccc}
        \toprule
        \textbf{Criterion} & \textbf{SIG} & \textbf{BAK} & \textbf{OVRL} & \textbf{NISQA} & \textbf{SBERT} & \textbf{SIM} \\
        \midrule
        - & 3.550 & 4.097 & 3.294 & 4.556 & 0.788 & 0.894 \\
        Multi-Metric & \textbf{3.564} & \textbf{4.144} & \textbf{3.331} & \textbf{4.850} & 0.803 & \textbf{0.904} \\
        NISQA & 3.531 & 4.137 & 3.300 & 4.810 & 0.785 & 0.896 \\
        OVRL & 3.561 & 4.117 & 3.317 & 4.600 & 0.792 & 0.896 \\
        SIM & 3.537 & 4.101 & 3.285 & 4.577 & 0.792 & 0.901 \\
        SBERT & 3.540 & 4.109 & 3.291 & 4.612 & \textbf{0.804} & 0.901 \\
        \bottomrule
    \end{tabular}
    \caption{Ablation on preference metric for the AR model on Voicefixer-GSR testset. The Multi-Metric Preference Strategy ("Multi-Metric") consistently outperforms single-metric approaches across all evaluation metrics.}
    \label{tab:metric_ablation}
\end{table}

\begin{figure}[!t]
    \centering
    \subfloat[SIG]{
        \includegraphics[width=0.48\linewidth]{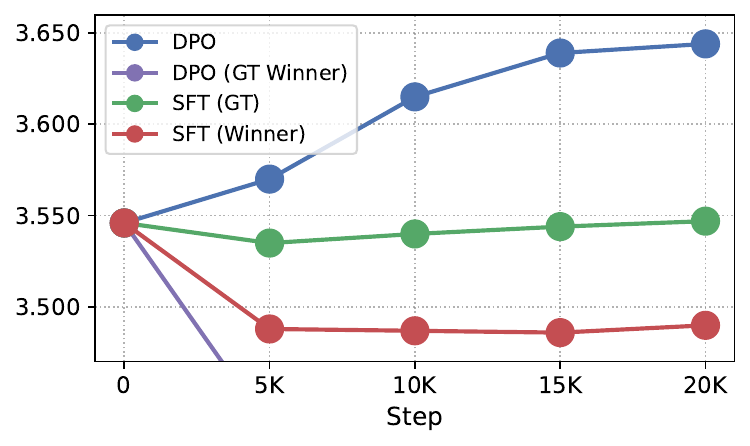}
        \label{fig:sig}
    }
    \subfloat[BAK]{
        \includegraphics[width=0.48\linewidth]{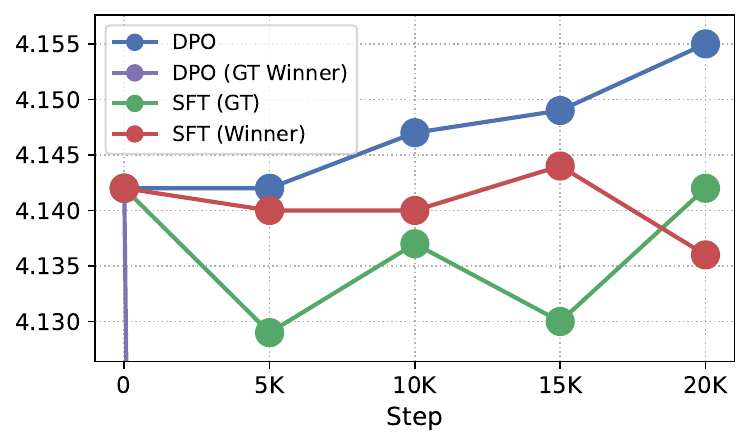}
        \label{fig:bak}
    }
    \hfill
    \subfloat[OVRL]{
        \includegraphics[width=0.48\linewidth]{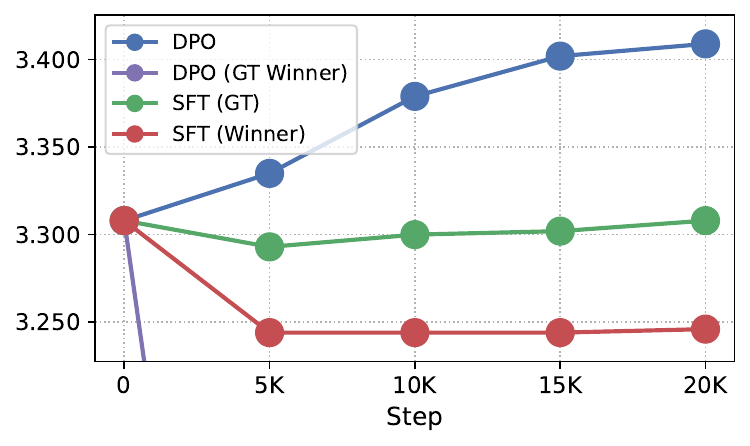}
        \label{fig:ovrl}
    }
    \subfloat[NISQA]{
        \includegraphics[width=0.48\linewidth]{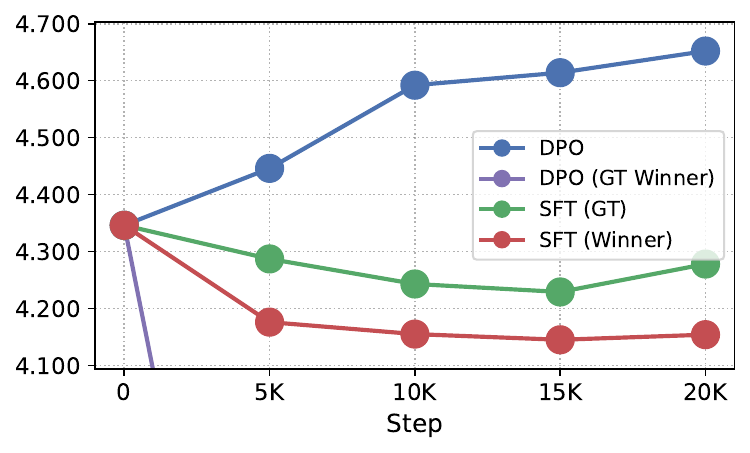}
        \label{fig:nisqa}
    }
    \hfill
    \subfloat[SpeechBERTScore]{
        \includegraphics[width=0.48\linewidth]{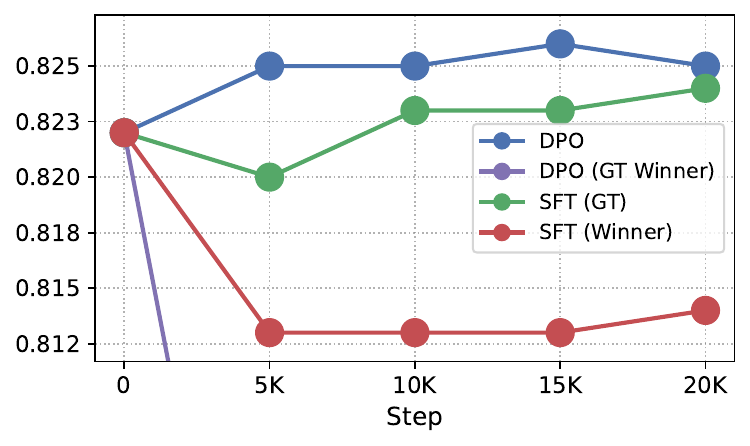}
        \label{fig:speechbertscore}
    }
    \subfloat[Similarity]{
        \includegraphics[width=0.48\linewidth]{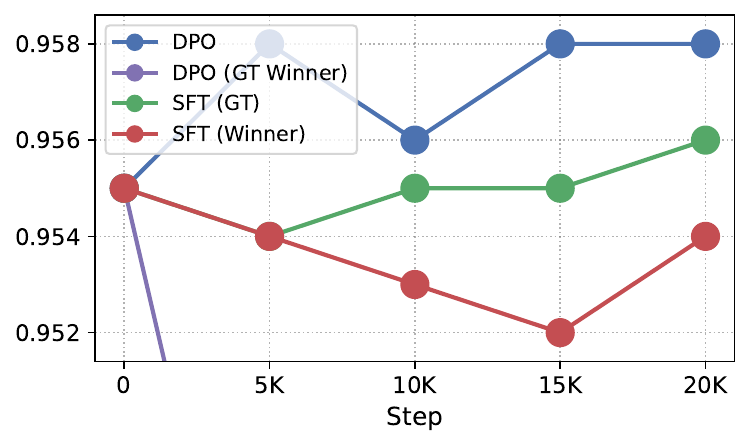}
        \label{fig:similarity}
    }
    
    \caption{Ablation study on training objectives for the MGM model. DPO demonstrates consistent improvements across training steps, while SFT tends to stagnate or degrade. Notably, a naive DPO variant that treats ground-truth outputs as winners ("GT Winner") results in model collapse.}
    \label{fig:ablation_sft}
\end{figure}

\begin{figure}[h]
    \centering
    \subfloat[Reward Margin ($\times 10^5$)]{
        \includegraphics[width=0.48\linewidth]{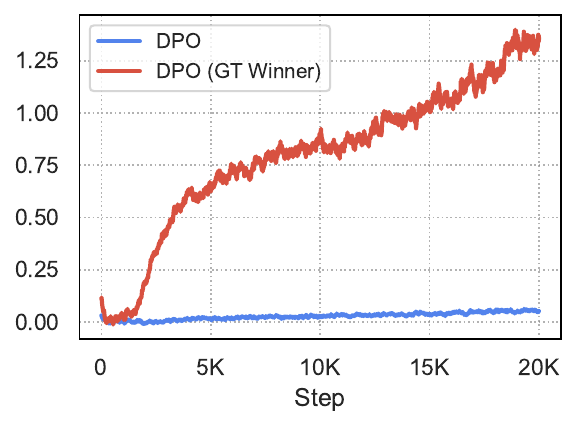}
        \label{fig:reward_margin}
    }
    \subfloat[Reward Accuracy]{
        \includegraphics[width=0.48\linewidth]{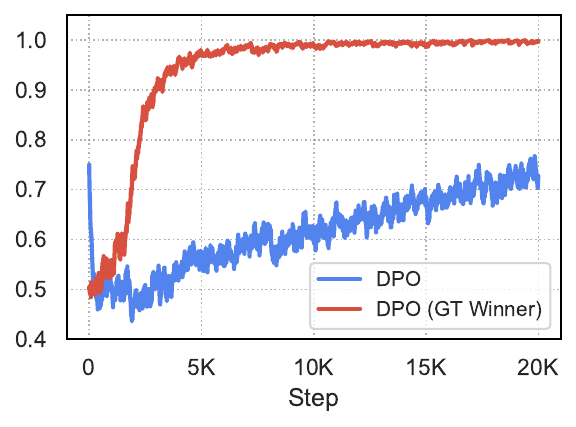}
        \label{fig:reward_accuracy}
    }
    
    \caption{Detailed training curves between normal DPO and DPO (GT Winner) training (smoothing factor = 0.99). Using ground truth as the unconditional winner leads to inflated reward margins and saturated reward accuracy, indicating model collapse.}
    \label{fig:loss_gt_winner}
\end{figure}

\noindent\textbf{DPO vs. Supervised Fine-Tuning (SFT).} To highlight the benefits of preference-based optimization, we compare DPO against two SFT baselines: fine-tuning on ground-truth audio (SFT-GT) and on the "winner" samples from our preference data (SFT-Winner). As shown in Figure~\ref{fig:ablation_sft}, DPO consistently outperforms both SFT variants, which show only marginal gains before stagnating. This suggests that simply exposing the model to high-quality examples is insufficient for effective alignment. 
Furthermore, a naive DPO strategy using ground-truth as the fixed ``winner'' leads to model collapse, as detailed in Figure~\ref{fig:loss_gt_winner}.
This behavior is driven by the core DPO objective, which seeks to maximize the log-probability ratio, $\beta \log \frac{p_\theta(y_0^w|y_t^w, x)}{p_{\text{ref}}(y_0^w|y_t^w, x)} - \beta \log \frac{p_\theta(y_0^l|y_t^l, x)}{p_{\text{ref}}(y_0^l|y_t^l, x)}$. When $y_w$ is a fixed ground-truth target, the model learns a pathological shortcut—drastically suppressing the probability of all outputs other than the GT to maximize this ratio. This highlights the necessity of learning from relative, nuanced preferences.

\begin{table}[ht]
    \centering
    \small
    \renewcommand{\arraystretch}{1}
    \setlength{\tabcolsep}{1mm}
    \begin{tabular}{lcccccc}
        \toprule
        \textbf{Model (Source)} & \textbf{SIG} & \textbf{BAK} & \textbf{OVRL} & \textbf{NISQA} & \textbf{SBERT} & \textbf{SIM} \\
        \midrule
        
        AR (-) & 3.663 & 4.134 & 3.419 & 4.535 & 0.783 & 0.922 \\
        AR (AR) & 3.672 & \textbf{4.171} & \textbf{3.449} & \textbf{4.867} & \textbf{0.791} & \textbf{0.925} \\
        AR (MGM) & \textbf{3.675} & 4.148 & 3.435 & 4.604 & 0.786 & 0.919 \\
        AR (FM) & 3.668 & 4.166 & 3.440 & 4.783 & 0.788 & 0.918 \\
        
        \midrule
        
        FM (-) & 3.550 & 4.062 & 3.281 & 4.184 & 0.791 & \textbf{0.931} \\
        FM (AR) & 3.575 & 4.103 & 3.324 & 4.473 & 0.794 & 0.932 \\
        FM (MGM) & \textbf{3.587} & 4.048 & 3.305 & 4.269 & 0.789 & 0.930 \\
        FM (FM) & 3.580 & \textbf{4.126} & \textbf{3.337} & \textbf{4.645} & \textbf{0.797} & 0.929 \\
        
        \midrule
        
        MGM (-) & 3.546 & 4.142 & 3.308 & 4.346 & 0.822 & 0.955 \\
        MGM (AR) & 3.589 & 4.147 & 3.346 & 4.321 & 0.809 & 0.951 \\
        MGM (MGM) & \textbf{3.644} & \textbf{4.155} & \textbf{3.409} & \textbf{4.652} & \textbf{0.825} & \textbf{0.958} \\
        MGM (FM) & 3.478 & 4.154 & 3.232 & 4.468 & 0.802 & 0.946 \\
        
        \bottomrule
    \end{tabular}
    \caption{Ablation on the source of preference data, evaluated on the Librivox-GSR testset. For each base model, performance is compared against its baseline (-) when aligned with data from different model paradigms (AR, MGM, FM).}
    \label{tab:ablation_source}
\end{table}

\begin{table}[h!]
    \centering
    \small
    \begin{tabular}{lccc}
    \toprule
     & \textbf{AR} & \textbf{FM} & \textbf{MGM} \\
    \midrule
    \textbf{AR}  & \textbf{0.0075} & 0.0072 & 0.0018 \\
    \textbf{FM}  & 0.0072 & \textbf{0.0102} & 0.0005 \\
    \textbf{MGM} & 0.0018 & 0.0005 & \textbf{0.0045} \\
    \bottomrule
    \end{tabular}
    \caption{Average pairwise cosine similarity of preference vectors ($v_{\text{pref}} = f(y_w) - f(y_l)$) within and across generative paradigms. Higher values indicate greater alignment in the `alignment direction'. The strong diagonal highlights that in-paradigm preference data is more internally consistent.}
    \label{tab:cosine_similarity}
\end{table}

\noindent\textbf{The Role of Data Source in Alignment.}
Finally, we explore how the source of preference data affects alignment performance. Table~\ref{tab:ablation_source} shows the results of aligning each model with preference pairs generated from itself (in-paradigm) versus from other paradigms (cross-paradigm). This distinction is analogous to the concepts of on-policy and off-policy learning in reinforcement learning. We observe a clear trend: all three models achieve their optimal performance when aligned with their own in-paradigm preference data. To quantitatively investigate this phenomenon, we formally define the principle of \textbf{in-paradigm alignment}. We hypothesize that each architecture possesses a unique ``preference direction''—an optimal path for improvement. To quantify this, we define a ``preference vector'' as the feature-space difference between winner and loser samples ($v_{\text{pref}} = f(y_w) - f(y_l)$) using pre-trained w2v-bert-2 model\footnote{\url{https://huggingface.co/facebook/w2v-bert-2.0}} and compute their average pairwise cosine similarity. The results (Table~\ref{tab:cosine_similarity}) provide strong quantitative evidence for our hypothesis: the average \textbf{in-paradigm similarity (diagonal values) is consistently higher than the cross-paradigm similarity} (e.g., AR-AR similarity of 0.0075 vs. AR-MGM of 0.0018). This indicates that data from a model's own paradigm provides a more consistent and targeted optimization signal. 

Furthermore, these similarities correlate with alignment performance (Table~\ref{tab:ablation_source}): paradigms with more aligned preference vectors (e.g., AR and FM, similarity 0.0072) show better cross-paradigm transfer performance than less aligned ones (e.g., FM and MGM, similarity 0.0005). This analysis provides strong backing for our principle, suggesting that alignment is most effective when the preference data's direction matches the target model's intrinsic generative manifold.

\subsection{Application: Empowering Discriminative Models via Pseudo-Labeling}
\label{sec:application}

\begin{table}[h]
    \centering
    \small
    \renewcommand{\arraystretch}{1}
    \begin{tabular}{lcccc}
    \toprule
    \textbf{Model} & \textbf{SIG} & \textbf{BAK} & \textbf{OVRL} & \textbf{NISQA} \\
    \midrule
    Voicefixer (before) & 2.657 & 3.080 & 2.295 & 2.919 \\
    Voicefixer (after) & \textbf{3.096} & \textbf{3.745} & \textbf{2.756} & \textbf{3.312} \\
    \bottomrule
    \end{tabular}
    \caption{Evaluation on real-world singing-voice recordings before and after DPO model's annotation}
    \label{tab:result-signal-model}
\end{table}

We demonstrate our aligned model's utility as a ``data annotator'' in a data-scarce scenario by fine-tuning the Voicefixer restoration model on singing recordings. Instead of ground-truth data, we used pseudo-labels generated by our DPO-enhanced \anyenhance model as supervision. This resulted in dramatic performance improvements across all metrics (Table~\ref{tab:result-signal-model}). Our findings show a powerful generative model can effectively create supervision signals to train smaller, discriminative models when paired data is prohibitively difficult to obtain. This bridges the gap between the two modeling paradigms and opens up new avenues for practical speech restoration. \todo{More detail on this application can be found in the Appendix~\ref{app:application_pseudo_labeling}.}

\section{Conclusion}

In this work, we introduced a \textbf{multi-metric preference alignment} strategy to align generative speech restoration models with human perception while mitigating the risk of reward hacking. To this end, we constructed a new dataset, \textbf{\tool}, using a strict unanimous agreement criterion across four complementary metrics to ensure a holistic and robust preference signal. Applying DPO with our dataset yields significant improvements across diverse generative paradigms (AR, MGM, and FM), outperforming single-metric preference and demonstrating the principle of in-paradigm alignment. Furthermore, we demonstrated a novel and practical application where our aligned models serve as powerful ``data annotators'', generating high-quality pseudo-labels to effectively train discriminative models in data-scarce scenarios. Future work includes exploring more advanced alignment algorithms, investigating preference data properties, and leveraging our annotation capabilities for large-scale data curation in other domains like TTS.

\section*{Acknowledgments}
This work is funded by the NSFC under Grant 62376237, Shenzhen Science and Technology Program ZDSYS20230626091302006, 2023 Shenzhen stability Science Program, Internal Project Fund from Shenzhen Research Institute of Big Data, Grant No. T00120230002, and Program for Guangdong Introducing Innovative and Enterpreneurial Teams, Grant No. 2023ZT10X044.

\bibliography{ref}

@article{vaswani2017attention,
  title={Attention is all you need},
  author={Vaswani, Ashish and Shazeer, Noam and Parmar, Niki and Uszkoreit, Jakob and Jones, Llion and Gomez, Aidan N and Kaiser, {\L}ukasz and Polosukhin, Illia},
  journal={NeurIPS},
  volume={30},
  year={2017}
}

@article{kingma2014adam,
  title={Adam: A method for stochastic optimization},
  author={Kingma, Diederik P and Ba, Jimmy},
  journal={arXiv preprint arXiv:1412.6980},
  year={2014}
}

@inproceedings{chang2022maskgit,
  title={Maskgit: Masked generative image transformer},
  author={Chang, Huiwen and Zhang, Han and Jiang, Lu and Liu, Ce and Freeman, William T},
  booktitle={CVPR},
  year={2022}
}

@inproceedings{wang2024maskgct,
  title={Mask{GCT}: Zero-Shot Text-to-Speech with Masked Generative Codec Transformer},
  author={Yuancheng Wang and Haoyue Zhan and Liwei Liu and Ruihong Zeng and Haotian Guo and Jiachen Zheng and Qiang Zhang and Xueyao Zhang and Shunsi Zhang and Zhizheng Wu},
  booktitle={ICLR},
  year={2025},
}

@article{borsos2023soundstorm,
  title={Soundstorm: Efficient parallel audio generation},
  author={Borsos, Zal{\'a}n and Sharifi, Matt and Vincent, Damien and Kharitonov, Eugene and Zeghidour, Neil and Tagliasacchi, Marco},
  journal={arXiv preprint arXiv:2305.09636},
  year={2023}
}

@article{wang2025metis,
  title={Metis: A Foundation Speech Generation Model with Masked Generative Pre-training},
  author={Wang, Yuancheng and Zheng, Jiachen and Zhang, Junan and Zhang, Xueyao and Liao, Huan and Wu, Zhizheng},
  journal={arXiv preprint arXiv:2502.03128},
  year={2025}
}

@inproceedings{ju2024naturalspeech,
  title={NaturalSpeech 3: zero-shot speech synthesis with factorized codec and diffusion models},
  author={Ju, Zeqian and Wang, Yuancheng and Shen, Kai and Tan, Xu and Xin, Detai and Yang, Dongchao and Liu, Yanqing and Leng, Yichong and Song, Kaitao and Tang, Siliang and others},
  booktitle={ICML},
  year={2024}
}

@article{lipman2022flow,
  title={Flow matching for generative modeling},
  author={Lipman, Yaron and Chen, Ricky TQ and Ben-Hamu, Heli and Nickel, Maximilian and Le, Matt},
  journal={arXiv preprint arXiv:2210.02747},
  year={2022}
}

@inproceedings{peebles2023scalable,
  title={Scalable diffusion models with transformers},
  author={Peebles, William and Xie, Saining},
  booktitle={ICCV},
  pages={4195--4205},
  year={2023}
}

@article{kumar2024high,
  title={High-fidelity audio compression with improved rvqgan},
  author={Kumar, Rithesh and Seetharaman, Prem and Luebs, Alejandro and Kumar, Ishaan and Kumar, Kundan},
  journal={NeurIPS},
  year={2024}
}

@article{li2024masksr,
  title={MaskSR: Masked Language Model for Full-band Speech Restoration},
  author={Li, Xu and Wang, Qirui and Liu, Xiaoyu},
  journal={arXiv preprint arXiv:2406.02092},
  year={2024}
}

@article{liu2024joint,
  title={Joint Semantic Knowledge Distillation and Masked Acoustic Modeling for Full-band Speech Restoration with Improved Intelligibility},
  author={Liu, Xiaoyu and Li, Xu and Serr{\`a}, Joan and Pascual, Santiago},
  journal={arXiv preprint arXiv:2409.09357},
  year={2024}
}

@inproceedings{wang2024selm,
  title={SELM: Speech enhancement using discrete tokens and language models},
  author={Wang, Ziqian and Zhu, Xinfa and Zhang, Zihan and Lv, YuanJun and Jiang, Ning and Zhao, Guoqing and Xie, Lei},
  booktitle={ICASSP},
  year={2024},
}

@inproceedings{yang2024genhancer,
  title={Genhancer: High-Fidelity Speech Enhancement via Generative Modeling on Discrete Codec Tokens},
  author={Yang, Haici and Su, Jiaqi and Kim, Minje and Jin, Zeyu},
  booktitle={INTERSPEECH},
  year={2024}
}

@article{liu2022voicefixer,
  title={Voicefixer: A unified framework for high-fidelity speech restoration},
  author={Liu, Haohe and Liu, Xubo and Kong, Qiuqiang and Tian, Qiao and Zhao, Yan and Wang, DeLiang and Huang, Chuanzeng and Wang, Yuxuan},
  journal={arXiv preprint arXiv:2204.05841},
  year={2022}
}

@inproceedings{liu2024audiosr,
  title={AudioSR: Versatile audio super-resolution at scale},
  author={Liu, Haohe and Chen, Ke and Tian, Qiao and Wang, Wenwu and Plumbley, Mark D},
  booktitle={ICASSP},
  year={2024},
}

@article{tang2024tselm,
  title={TSELM: Target Speaker Extraction using Discrete Tokens and Language Models},
  author={Tang, Beilong and Zeng, Bang and Li, Ming},
  journal={arXiv preprint arXiv:2409.07841},
  year={2024}
}

@inproceedings{reddy2020interspeech,
  title={The INTERSPEECH 2020 deep noise suppression challenge: Datasets, subjective testing framework, and challenge results},
  author={Reddy, Chandan KA and Gopal, Vishak and Cutler, Ross and Beyrami, Ebrahim and Cheng, Roger and Dubey, Harishchandra and Matusevych, Sergiy and Aichner, Robert and Aazami, Ashkan and Braun, Sebastian and others},
  booktitle={INTERSPEECH},
  year={2020}
}

@inproceedings{zhang24h_interspeech,
  title     = {URGENT Challenge: Universality, Robustness, and Generalizability For Speech Enhancement},
  author    = {Wangyou Zhang and Robin Scheibler and Kohei Saijo and Samuele Cornell and Chenda Li and Zhaoheng Ni and Jan Pirklbauer and Marvin Sach and Shinji Watanabe and Tim Fingscheidt and Yanmin Qian},
  year      = {2024},
  booktitle = {INTERSPEECH},
  doi       = {10.21437/Interspeech.2024-1239},
}

@misc{ccmusic,
  author       = {Zhou, Monan and Xu, Shenyang and Liu, Zhaorui and Wang, Zhaowen and Yu, Feng and Li, Wei and Han, Baoqiang},
  title        = {CCMusic: An Open and Diverse Database for Chinese and General Music Information Retrieval Research},
  year         = {2024},
  publisher    = {HuggingFace},
}

@article{he2024emilia,
  title={Emilia: An extensive, multilingual, and diverse speech dataset for large-scale speech generation},
  author={He, Haorui and Shang, Zengqiang and Wang, Chaoren and Li, Xuyuan and Gu, Yicheng and Hua, Hua and Liu, Liwei and Yang, Chen and Li, Jiaqi and Shi, Peiyang and others},
  journal={arXiv preprint arXiv:2407.05361},
  year={2024}
}

@article{kearns2014librivox,
  title={Librivox: Free public domain audiobooks},
  author={Kearns, Jodi},
  journal={Reference Reviews},
  year={2014},
}

@inproceedings{reddy2022dnsmos,
  title={DNSMOS P.835: A non-intrusive perceptual objective speech quality metric to evaluate noise suppressors},
  author={Reddy, Chandan KA and Gopal, Vishak and Cutler, Ross},
  booktitle={ICASSP},
  year={2022}
}

@inproceedings{mittag2021nisqa,
  author       = {Gabriel Mittag and Babak Naderi and Assmaa Chehadi and Sebastian M{\"{o}}ller},
  title        = {NISQA: A Deep CNN-Self-Attention Model for Multidimensional Speech Quality Prediction with Crowdsourced Datasets},
  booktitle    = {INTERSPEECH},
  year         = {2021},
  doi          = {10.21437/INTERSPEECH.2021-299},
}

@article{saeki2024speechbertscore,
  title={SpeechBERTScore: Reference-Aware Automatic Evaluation of Speech Generation Leveraging NLP Evaluation Metrics},
  author={Saeki, Takaaki and Maiti, Soumi and Takamichi, Shinnosuke and Watanabe, Shinji and Saruwatari, Hiroshi},
  journal={arXiv preprint arXiv:2401.16812},
  year={2024}
}

@inproceedings{liugenerative,
  title={Generative Pre-training for Speech with Flow Matching},
  author={Liu, Alexander H and Le, Matthew and Vyas, Apoorv and Shi, Bowen and Tjandra, Andros and Hsu, Wei-Ning},
  booktitle={ICLR},
  year={2024}
}

@inproceedings{Yang2024UniAudio,
  author       = {Dongchao Yang and Jinchuan Tian and Xu Tan and others},
  title        = {UniAudio: Towards Universal Audio Generation with Large Language Models},
  booktitle    = {ICML},
  year         = {2024},
}

@article{ku2024generative,
  title={Generative Speech Foundation Model Pretraining for High-Quality Speech Extraction and Restoration},
  author={Ku, Pin-Jui and Liu, Alexander H and Korostik, Roman and Huang, Sung-Feng and Fu, Szu-Wei and Juki{\'c}, Ante},
  journal={arXiv preprint arXiv:2409.16117},
  year={2024}
}

@article{wang2024speechx,
  title={Speechx: Neural codec language model as a versatile speech transformer},
  author={Wang, Xiaofei and Thakker, Manthan and Chen, Zhuo and Kanda, Naoyuki and Eskimez, Sefik Emre and Chen, Sanyuan and Tang, Min and Liu, Shujie and Li, Jinyu and Yoshioka, Takuya},
  journal={IEEE/ACM Transactions on Audio, Speech, and Language Processing},
  year={2024},
}

@inproceedings{anonymous2025gense,
  title={Gen{SE}: Generative Speech Enhancement via Language Models using Hierarchical Modeling},
  author={Jixun Yao and Hexin Liu and Chen Chen and Yuchen Hu and EngSiong Chng and Lei Xie},
  booktitle={ICLR},
  year={2025},
}

@article{kang2025llase,
  title={LLaSE-G1: Incentivizing generalization capability for llama-based speech enhancement},
  author={Kang, Boyi and Zhu, Xinfa and Zhang, Zihan and Ye, Zhen and Liu, Mingshuai and Wang, Ziqian and Zhu, Yike and Ma, Guobin and Chen, Jun and Xiao, Longshuai and others},
  journal={arXiv preprint arXiv:2503.00493},
  year={2025}
}

@article{wang2025flowse,
  title={FlowSE: Efficient and High-Quality Speech Enhancement via Flow Matching},
  author={Wang, Ziqian and Liu, Zikai and Zhu, Xinfa and Zhu, Yike and Liu, Mingshuai and Chen, Jun and Xiao, Longshuai and Weng, Chao and Xie, Lei},
  journal={arXiv preprint arXiv:2505.19476},
  year={2025}
}

@article{zhang2025anyenhance,
  title={AnyEnhance: A Unified Generative Model With Prompt-Guidance and Self-Critic for Voice Enhancement},
  author={Zhang, Junan and Yang, Jing and Fang, Zihao and Wang, Yuancheng and Zhang, Zehua and Wang, Zhuo and Fan, Fan and Wu, Zhizheng},
  journal={IEEE Transactions on Audio, Speech and Language Processing},
  year={2025},
}

@article{rafailov2023direct,
  title={Direct preference optimization: Your language model is secretly a reward model},
  author={Rafailov, Rafael and Sharma, Archit and Mitchell, Eric and Manning, Christopher D and Ermon, Stefano and Finn, Chelsea},
  journal={NeurIPS},
  volume={36},
  pages={53728--53741},
  year={2023}
}

@inproceedings{zhang2025advancing,
  title={Advancing zero-shot text-to-speech intelligibility across diverse domains via preference alignment},
  author={Zhang, Xueyao and Wang, Yuancheng and Wang, Chaoren and Li, Ziniu and Chen, Zhuo and Wu, Zhizheng},
  booktitle={ACL},
  pages={12251--12270},
  year={2025}
}

@article{zhang2025vevo,
  title={Vevo: Controllable zero-shot voice imitation with self-supervised disentanglement},
  author={Zhang, Xueyao and Zhang, Xiaohui and Peng, Kainan and Tang, Zhenyu and Manohar, Vimal and Liu, Yingru and Hwang, Jeff and Li, Dangna and Wang, Yuhao and Chan, Julian and others},
  journal={arXiv preprint arXiv:2502.07243},
  year={2025}
}

@article{siuzdak2023vocos,
  title={Vocos: Closing the gap between time-domain and fourier-based neural vocoders for high-quality audio synthesis},
  author={Siuzdak, Hubert},
  journal={arXiv preprint arXiv:2306.00814},
  year={2023}
}

@article{sun2025f5r,
  title={F5R-TTS: Improving flow-matching based text-to-speech with group relative policy optimization},
  author={Sun, Xiaohui and Xiao, Ruitong and Mo, Jianye and Wu, Bowen and Yu, Qun and Wang, Baoxun},
  journal={arXiv preprint arXiv:2504.02407},
  year={2025}
}

@article{hussain2025koel,
  title={Koel-TTS: Enhancing LLM based Speech Generation with Preference Alignment and Classifier Free Guidance},
  author={Hussain, Shehzeen and Neekhara, Paarth and Yang, Xuesong and Casanova, Edresson and Ghosh, Subhankar and Desta, Mikyas T and Fejgin, Roy and Valle, Rafael and Li, Jason},
  journal={arXiv preprint arXiv:2502.05236},
  year={2025}
}

@inproceedings{tian2025preference,
  title={Preference alignment improves language model-based tts},
  author={Tian, Jinchuan and Zhang, Chunlei and Shi, Jiatong and Zhang, Hao and Yu, Jianwei and Watanabe, Shinji and Yu, Dong},
  booktitle={ICASSP},
  pages={1--5},
  year={2025},
  organization={IEEE}
}

@article{yao2025fine,
  title={Fine-grained Preference Optimization Improves Zero-shot Text-to-Speech},
  author={Yao, Jixun and Yang, Yuguang and Pan, Yu and Feng, Yuan and Ning, Ziqian and Ye, Jianhao and Zhou, Hongbin and Xie, Lei},
  journal={arXiv preprint arXiv:2502.02950},
  year={2025}
}

@inproceedings{zhang2024speechalign,
  title={SpeechAlign: Aligning Speech Generation to Human Preferences},
  author={Zhang, Dong and Li, Zhaowei and Li, Shimin and Zhang, Xin and Wang, Pengyu and Zhou, Yaqian and Qiu, Xipeng},
  booktitle={NeurIPS},
  year={2024}
}

@article{gu2025singnet,
  title={Singnet: Towards a large-scale, diverse, and in-the-wild singing voice dataset},
  author={Gu, Yicheng and Wang, Chaoren and Zhang, Junan and Zhang, Xueyao and Fang, Zihao and He, Haorui and Wu, Zhizheng},
  journal={arXiv preprint arXiv:2505.09325},
  year={2025}
}

@article{amphion_v0.2,
  title        = {Overview of the Amphion Toolkit (v0.2)},
  author       = {Jiaqi Li and Xueyao Zhang and Yuancheng Wang and Haorui He and Chaoren Wang and Li Wang and Huan Liao and Junyi Ao and Zeyu Xie and Yiqiao Huang and Junan Zhang and Zhizheng Wu},
  journal      = {arXiv preprint arXiv:2501.15442},
  year         = {2025},
}

@inproceedings{amphion,
  author={Zhang, Xueyao and Xue, Liumeng and Gu, Yicheng and Wang, Yuancheng and Li, Jiaqi and He, Haorui and Wang, Chaoren and Song, Ting and Chen, Xi and Fang, Zihao and Chen, Haopeng and Zhang, Junan and Tang, Tze Ying and Zou, Lexiao and Wang, Mingxuan and Han, Jun and Chen, Kai and Li, Haizhou and Wu, Zhizheng},
  title={Amphion: An Open-Source Audio, Music and Speech Generation Toolkit},
  booktitle={SLT},
  year={2024},
}

@inproceedings{fu2019metricgan,
  title={Metricgan: Generative adversarial networks based black-box metric scores optimization for speech enhancement},
  author={Fu, Szu-Wei and Liao, Chien-Feng and Tsao, Yu and Lin, Shou-De},
  booktitle={ICML},
  pages={2031--2041},
  year={2019}
}

@inproceedings{kumar2025using,
  title={Using RLHF to align speech enhancement approaches to mean-opinion quality scores},
  author={Kumar, Anurag and Perrault, Andrew and Williamson, Donald S},
  booktitle={ICASSP},
  pages={1--5},
  year={2025}
}

@article{li2025aligning,
  title={Aligning Generative Speech Enhancement with Human Preferences via Direct Preference Optimization},
  author={Li, Haoyang and Hou, Nana and Hu, Yuchen and Yao, Jixun and Siniscalchi, Sabato Marco and Chng, Eng Siong},
  journal={arXiv preprint arXiv:2507.09929},
  year={2025}
}

@article{welker2022sgmse,
  title={Speech enhancement with score-based generative models in the complex STFT domain},
  author={Welker, Simon and Richter, Julius and Gerkmann, Timo},
  journal={arXiv preprint arXiv:2203.17004},
  year={2022}
}

@inproceedings{lemercier2023analysing,
  title={Analysing diffusion-based generative approaches versus discriminative approaches for speech restoration},
  author={Lemercier, Jean-Marie and Richter, Julius and Welker, Simon and Gerkmann, Timo},
  booktitle={ICASSP},
  pages={1--5},
  year={2023}
}

@inproceedings{li2025bridgesr,
  title={Bridge-sr: Schr{\"o}dinger bridge for efficient sr},
  author={Li, Chang and Chen, Zehua and Bao, Fan and Zhu, Jun},
  booktitle={ICASSP},
  pages={1--5},
  year={2025}
}

@article{jukic2024bridgese,
  title={Schr{\"o}dinger bridge for Generative Speech Enhancement},
  author={Juki{\'c}, Ante and Korostik, Roman and Balam, Jagadeesh and Ginsburg, Boris},
  journal={arXiv preprint arXiv:2407.16074},
  year={2024}
}

@article{wang2025solospeech,
  title={SoloSpeech: Enhancing Intelligibility and Quality in Target Speech Extraction through a Cascaded Generative Pipeline},
  author={Wang, Helin and Hai, Jiarui and Yang, Dongchao and Chen, Chen and Li, Kai and Peng, Junyi and Thebaud, Thomas and Velazquez, Laureano Moro and Villalba, Jesus and Dehak, Najim},
  journal={arXiv preprint arXiv:2505.19314},
  year={2025}
}

@article{liu2025videoalign,
  title={Improving video generation with human feedback},
  author={Liu, Jie and Liu, Gongye and Liang, Jiajun and Yuan, Ziyang and Liu, Xiaokun and Zheng, Mingwu and Wu, Xiele and Wang, Qiulin and Qin, Wenyu and Xia, Menghan and others},
  journal={arXiv preprint arXiv:2501.13918},
  year={2025}
}

@article{ouyang2022instructgpt,
  title={Training language models to follow instructions with human feedback},
  author={Ouyang, Long and Wu, Jeffrey and Jiang, Xu and Almeida, Diogo and Wainwright, Carroll and Mishkin, Pamela and Zhang, Chong and Agarwal, Sandhini and Slama, Katarina and Ray, Alex and others},
  journal={NeurIPS},
  volume={35},
  pages={27730--27744},
  year={2022}
}

@article{bai2022rlhf,
  title={Training a helpful and harmless assistant with reinforcement learning from human feedback},
  author={Bai, Yuntao and Jones, Andy and Ndousse, Kamal and Askell, Amanda and Chen, Anna and DasSarma, Nova and Drain, Dawn and Fort, Stanislav and Ganguli, Deep and Henighan, Tom and others},
  journal={arXiv preprint arXiv:2204.05862},
  year={2022}
}

@article{xu2023imagereward,
  title={Imagereward: Learning and evaluating human preferences for text-to-image generation},
  author={Xu, Jiazheng and Liu, Xiao and Wu, Yuchen and Tong, Yuxuan and Li, Qinkai and Ding, Ming and Tang, Jie and Dong, Yuxiao},
  journal={NeurIPS},
  volume={36},
  pages={15903--15935},
  year={2023}
}

@inproceedings{wallace2024diffusiondpo,
  title={Diffusion model alignment using direct preference optimization},
  author={Wallace, Bram and Dang, Meihua and Rafailov, Rafael and Zhou, Linqi and Lou, Aaron and Purushwalkam, Senthil and Ermon, Stefano and Xiong, Caiming and Joty, Shafiq and Naik, Nikhil},
  booktitle={CVPR},
  pages={8228--8238},
  year={2024}
}

@inproceedings{cideron2024musicrl,
  title={MusicRL: Aligning Music Generation to Human Preferences},
  author={Cideron, Geoffrey and Girgin, Sertan and Verzetti, Mauro and Vincent, Damien and Kastelic, Matej and Borsos, Zal{\'a}n and Mcwilliams, Brian and Ungureanu, Victor and Bachem, Olivier and Pietquin, Olivier and others},
  booktitle={ICML},
  pages={8968--8984},
  year={2024},
  organization={PMLR}
}

@article{lei2025levo,
  title={LeVo: High-Quality Song Generation with Multi-Preference Alignment},
  author={Lei, Shun and Xu, Yaoxun and Lin, Zhiwei and Zhang, Huaicheng and Tan, Wei and Chen, Hangting and Yu, Jianwei and Zhang, Yixuan and Yang, Chenyu and Zhu, Haina and others},
  journal={arXiv preprint arXiv:2506.07520},
  year={2025}
}

@inproceedings{majumder2024tango2,
  title={Tango 2: Aligning diffusion-based text-to-audio generations through direct preference optimization},
  author={Majumder, Navonil and Hung, Chia-Yu and Ghosal, Deepanway and Hsu, Wei-Ning and Mihalcea, Rada and Poria, Soujanya},
  booktitle={ACM Multimedia},
  pages={564--572},
  year={2024}
}

@article{liao2024baton,
  title={Baton: Aligning text-to-audio model with human preference feedback},
  author={Liao, Huan and Han, Haonan and Yang, Kai and Du, Tianjiao and Yang, Rui and Xu, Zunnan and Xu, Qinmei and Liu, Jingquan and Lu, Jiasheng and Li, Xiu},
  journal={arXiv preprint arXiv:2402.00744},
  year={2024}
}

\appendix



\section{Derivation of DPO Loss for Three Generative Paradigms}
\label{app:dpo_derivation}

This section provides a detailed derivation of the Direct Preference Optimization (DPO) loss functions for the three generative paradigms explored in our work: Autoregressive (AR), Masked Generative Models (MGM), and Flow-Matching (FM). We begin with a summary table (Table~\ref{tab:dpo_paradigms_summary}) that juxtaposes the key components of DPO across these paradigms, followed by step-by-step derivations for each.

\begin{table*}[!htbp]
    \centering
    \scriptsize
    \begin{tabular}{
        p{0.05\textwidth} 
        >{\raggedright\arraybackslash}p{0.25\textwidth} 
        p{0.22\textwidth} 
        >{\raggedright\arraybackslash}p{0.38\textwidth}
    }
        \toprule
        \textbf{Paradigm} & \textbf{RL Objective (with KL Regularization)} & \textbf{Implicit Reward Mapping} & \textbf{Final DPO Loss Formulation} \\
        \midrule

        \textbf{AR} & 
        $\max_{p_\theta} \, \mathbb{E}_{y \sim p_\theta(y|x)}[r(x, y)] - \beta D_{\text{KL}}\left[ p_\theta(y|x) \,\|\, p_{\text{ref}}(y|x) \right]$ &

        $\begin{aligned}[t]
            r(x, y) &= \beta \log \frac{p^*_\theta(y|x)}{p_{\text{ref}}(y|x)} \\
            &\quad + \beta \log Z(x)
        \end{aligned}$ &

        $\mathcal{L}_{\text{DPO}} = - \mathbb{E}_{\mathcal{D}} \left[ \log \sigma \left(
            \beta \log \frac{p_\theta(y_w|x)}{p_{\text{ref}}(y_w|x)} - 
            \beta \log \frac{p_\theta(y_l|x)}{p_{\text{ref}}(y_l|x)} \right) \right]$ \\

        \midrule

        \textbf{MGM} & 
        $\max_{p_\theta} \, \mathbb{E}_{y_0, t, x}[r(y_0, x)] - \beta D_{\text{KL}}\left[ p_\theta(y_0|y_t, x) \,\|\, p_{\text{ref}}(y_0|y_t, x) \right]$ &

        $\begin{aligned}[t]
            r(y_0, x) &= \beta \log \frac{p^*_\theta(y_0|y_t, x)}{p_{\text{ref}}(y_0|y_t, x)} \\
            &\quad + \beta \log Z(y_t, x)
        \end{aligned}$ &

        $\begin{aligned}[t]
            \mathcal{L}_{\text{DPO-MGM}} = - \mathbb{E}_{\mathcal{D}, t} \left[ \log \sigma \left(
            \beta \log \frac{p_\theta(y_0^w|y_t^w, x)}{p_{\text{ref}}(y_0^w|y_t^w, x)} \right. \right. \\
            \left. \left. \quad - \beta \log \frac{p_\theta(y_0^l|y_t^l, x)}{p_{\text{ref}}(y_0^l|y_t^l, x)} \right) \right]
        \end{aligned}$ \\

        \midrule

        \textbf{FM} & 
        $\max_{p_\theta} \, \mathbb{E}_{y_1, t, x}[r(y_1, x)] - \beta D_{\text{KL}}\left[ p_\theta(y_1|y_t, t, x) \,\|\, p_{\text{ref}}(y_1|y_t, t, x) \right]$ &

        $\begin{aligned}[t]
            r(y_1, x) &= \beta \log \frac{p^*_\theta(y_1|y_t, t, x)}{p_{\text{ref}}(y_1|y_t, t, x)} \\
            &\quad + \beta \log Z(y_t, t, x)
        \end{aligned}$ &

        $\begin{aligned}[t]
            \mathcal{L}_{\text{DPO-FM}} = - \mathbb{E}_{\mathcal{D}, t} \log \sigma\left(-\beta (\Delta_w - \Delta_l)\right), \\
            \text{where} \quad \Delta = \|v_\theta(y_t, t, x) - (y_1 - y_0)\|_2^2 \\
            \quad - \|v_{\text{ref}}(y_t, t, x) - (y_1 - y_0)\|_2^2
        \end{aligned}$ \\

        \bottomrule
    \end{tabular}
    \caption{Direct Preference Optimization (DPO) across different generative paradigms. The DPO framework is adapted from standard Autoregressive (AR) models to Masked Generative Models (MGM) and Flow-Matching (FM) models. This table summarizes the RL objective, implicit reward mapping, and final DPO loss formulation.}
    \label{tab:dpo_paradigms_summary}
\end{table*}

\subsection{DPO for Autoregressive Models}

The derivation for AR models follows the original DPO framework~\cite{rafailov2023direct}. The goal is to optimize a policy $p_\theta(y|x)$ to maximize a reward function $r(x,y)$, while regularizing its deviation from a reference policy $p_{\text{ref}}(y|x)$ using a KL-divergence term. The RL objective is:
\begin{equation}
    \max_{p_\theta} \, \mathbb{E}_{y \sim p_\theta(y|x)}[r(x, y)] - \beta D_{\text{KL}}\left[ p_\theta(y|x) \,\|\, p_{\text{ref}}(y|x) \right].
\end{equation}

The optimal policy that maximizes the RL objective is:

\begin{equation}
    p^*_\theta(y|x) = \frac{1}{Z(x)} p_{\text{ref}}(y|x) \exp\left( \frac{1}{\beta} r(x, y) \right),
\end{equation}

where $Z(x)$ ensures normalization. Thus, the reward can be written as:

\begin{equation}
    r(x, y) = \beta \log \frac{p^*_\theta(y|x)}{p_{\text{ref}}(y|x)} + \beta \log Z(x).
\end{equation}

DPO re-parameterizes the reward model using this relationship and substitutes it into the binary cross-entropy loss for a preference dataset $\mathcal{D} = \{(x, y_w, y_l)\}$, where $y_w$ is preferred over $y_l$. This results in a simple maximum likelihood objective that directly optimizes the policy $p_\theta$:

\begin{equation}
    \begin{aligned}
        \mathcal{L}_{\text{DPO}} = - \mathbb{E}_{(x, y_w, y_l) \sim \mathcal{D}} \Bigg[ \log \sigma \Big(
            \beta \log \frac{p_\theta(y_w|x)}{p_{\text{ref}}(y_w|x)} 
            {} \\
            - \beta \log \frac{p_\theta(y_l|x)}{p_{\text{ref}}(y_l|x)} \Big) \Bigg]
    \end{aligned}
\end{equation}

\subsection{DPO for Masked Generative Models (MGM)}

The RL objective is similar to that of AR models, but the policy to optimize is conditioned on a masked input \(y_t\) and a context \(x\):

\begin{equation}
    \begin{aligned}
    \max_{p_\theta} \; & \mathbb{E}_{y_0 \sim p_\theta(y_0|x),\, t,\, x} \left[ r(y_0, x) \right] \\
    & {} - \beta \, \text{D}_{\text{KL}} \left[ p_\theta(y_0|y_t, x) \, \| \, p_{\text{ref}}(y_0|y_t, x) \right].
    \end{aligned}
\end{equation}

The optimal policy takes the form:

\begin{equation}
    p^*_\theta(y_0|y_t, x) = \frac{1}{Z(y_t, x)} p_{\text{ref}}(y_0|y_t, x) \exp\left( \frac{1}{\beta} r(y_0, x) \right),
\end{equation}

and the reward becomes:

\begin{equation}
    r(y_0, x) = \beta \log \frac{p^*_\theta(y_0|y_t, x)}{p_{\text{ref}}(y_0|y_t, x)} + \beta \log Z(y_t, x).
\end{equation}

Substituting yields the DPO loss for MGM:
\begin{equation}
    \begin{aligned}
        \mathcal{L}_{\text{DPO-MGM}} = - \mathbb{E}_{(x, y_w, y_l) \sim \mathcal{D},\, t} \Bigg[ \log \sigma \Big(
        \beta \log \frac{p_\theta(y_0^w|y_t^w, x)}{p_{\text{ref}}(y_0^w|y_t^w, x)} \\
        {} - \beta \log \frac{p_\theta(y_0^l|y_t^l, x)}{p_{\text{ref}}(y_0^l|y_t^l, x)} \Big) \Bigg]
    \end{aligned}
\end{equation}

\subsection{DPO for Flow-Matching Models}

For Flow-Matching (FM) models, the RL objective is structured similarly while incorporating the flow-matching context \(y_t\) and time step \(t\) and the input \(x\):

\begin{equation}
    \begin{aligned}
    \max_{p_\theta} \; & \mathbb{E}_{y_1 \sim p_\theta(y_1|x), t, x} \left[ r(y_1, x) \right] \\
    & {} - \beta \, \text{D}_{\text{KL}} \left[ p_\theta(y_1|y_t, t, x) \, \| \, p_{\text{ref}}(y_1|y_t, t, x) \right].
    \end{aligned}
\end{equation}

The optimal policy:
\begin{equation}
    p^*_\theta(y_1|y_t, t, x) = \frac{1}{Z(y_t, t, x)} p_{\text{ref}}(y_1|y_t, t, x) \exp\left( \frac{1}{\beta} r(y_1, x) \right),
\end{equation}
with reward:
\begin{equation}
    r(y_1, x) = \beta \log \frac{p^*_\theta(y_1|y_t, t, x)}{p_{\text{ref}}(y_1|y_t, t, x)} + \beta \log Z(y_t, t, x).
\end{equation}

Because FM training assumes a Gaussian likelihood, following practice in~\cite{wallace2024diffusiondpo,liu2025videoalign}, we can express the policies into velocity predictions \(v_\theta\) and \(v_{\text{ref}}\) as follows:

\begin{equation}
    \begin{aligned}
        p_\theta(y_1|y_t, t, x) &\propto \exp\left( -\frac{1}{\beta} \| v_\theta(y_t, t, x) - (y_1 - y_0) \|_2^2 \right), \\
        p_{\text{ref}}(y_1|y_t, t, x) &\propto \exp\left( -\frac{1}{\beta} \| v_{\text{ref}}(y_t, t, x) - (y_1 - y_0) \|_2^2 \right).
    \end{aligned}
\end{equation}

Then we have the log-ratio:

\begin{equation}
    \begin{aligned}
    \log \frac{p_\theta(y_1|y_t, t, x)}{p_{\text{ref}}(y_1|y_t, t, x)} 
    = -\frac{1}{\beta} \Big(& \| v_\theta(y_t, t, x) - (y_1 - y_0) \|_2^2 \\
    & - \| v_{\text{ref}}(y_t, t, x) - (y_1 - y_0) \|_2^2 \Big).
    \end{aligned}
\end{equation}

Plugging into the DPO loss gives:

\begin{equation}
    \mathcal{L}_{\text{DPO-FM}} = - \mathbb{E}_{(x, y_w, y_l) \sim \mathcal{D}, t, y_0} \left[ \log \sigma \left( -\beta (\Delta_w - \Delta_l) \right) \right],
\end{equation}

where

\begin{equation}
    \Delta_w = \|v_\theta(y_t^w, t, x) - (y_1^w - y_0)\|_2^2 - \|v_{\text{ref}}(y_t^w, t, x) - (y_1^w - y_0)\|_2^2
\end{equation}

and $\Delta_l$ is defined analogously for the losing sample $y_l$. This loss encourages the model to reduce its velocity prediction error for preferred samples (making $\Delta_w$ more negative) while allowing for a larger error on dispreferred samples (making $\Delta_l$ less negative or positive) relative to the reference model.

\begin{table*}[htbp]
    \centering
    \small
    \setlength{\tabcolsep}{1mm}
    \begin{tabular}{cllcccccccc}
        \toprule
        \textbf{Dataset} & \textbf{Model} & \textbf{Type} & \makecell{\textbf{DPO-}\\\textbf{aligned?}} & \textbf{SIG$\uparrow$} & \textbf{BAK$\uparrow$} & \textbf{OVRL$\uparrow$} & \textbf{NISQA$\uparrow$} & \makecell{\textbf{Speech-}\\\textbf{BERTScore$\uparrow$}} & \textbf{Similarity$\uparrow$} \\
        \midrule
        
        \multirow{6}{*}{DNS-No-Reverb}
        & \multirow{2}{*}{\anyenhance} & \multirow{2}{*}{MGM} & \ding{55} & 3.640 & 4.179 & 3.418 & 4.821 & \textbf{0.907} & \textbf{0.988} \\
        &&& \ding{51} & \textbf{3.684} & \textbf{4.203} & \textbf{3.476} & \textbf{5.011} & 0.904 & 0.987 \\
        & \multirow{2}{*}{\arse} & \multirow{2}{*}{AR} & \ding{55} & 3.648 & 4.155 & 3.422 & 4.743 & 0.866 & \textbf{0.971} \\
        &&& \ding{51} & \textbf{3.673} & \textbf{4.194} & \textbf{3.469} & \textbf{4.974} & \textbf{0.868} & 0.970 \\
        & \multirow{2}{*}{\flowse} & \multirow{2}{*}{FM} & \ding{55} & 3.581 & 4.133 & 3.355 & 4.711 & 0.893 & \textbf{0.984} \\
        &&& \ding{51} & \textbf{3.632} & \textbf{4.173} & \textbf{3.420} & \textbf{4.998} & \textbf{0.895} & 0.983 \\
        
        \midrule
        
        \multirow{6}{*}{DNS-With-Reverb}
        & \multirow{2}{*}{\anyenhance} & \multirow{2}{*}{MGM} & \ding{55} & 3.500 & 4.040 & 3.204 & 3.722 & \textbf{0.738} & 0.951 \\
        &&& \ding{51} & \textbf{3.670} & \textbf{4.178} & \textbf{3.438} & \textbf{4.624} & 0.733 & \textbf{0.957} \\
        & \multirow{2}{*}{\arse} & \multirow{2}{*}{AR} & \ding{55} & 3.681 & 4.127 & 3.431 & 4.572 & \textbf{0.726} & \textbf{0.943} \\
        &&& \ding{51} & \textbf{3.709} & \textbf{4.189} & \textbf{3.496} & \textbf{4.957} & 0.715 & 0.940 \\
        & \multirow{2}{*}{\flowse} & \multirow{2}{*}{FM} & \ding{55} & 3.539 & 4.019 & 3.255 & 4.215 & 0.744 & 0.955 \\
        &&& \ding{51} & \textbf{3.629} & \textbf{4.163} & \textbf{3.399} & \textbf{4.924} & \textbf{0.746} & \textbf{0.955} \\
        
        \bottomrule
    \end{tabular}
    \caption{Quantitative comparison on DNS benchmarks across three generative paradigms before and after DPO alignment.}
    \label{tab:results-se}
\end{table*}

\section{Implementation Details}
\label{app:implementation_details}

\subsection{Model Details}

\noindent\textbf{Masked Generative Model (MGM).} We use \textbf{\anyenhance} as our model for MGM paradigm. \anyenhance~\cite{zhang2025anyenhance} is a unified masked generative model for speech and singing voice restoration, capable of handling a wide range of tasks (e.g., denoising, dereverberation, super-resolution) via prompt-guided in-context learning and a self-critic sampling mechanism. It is based on a transformer architecture~\cite{vaswani2017attention} trained with a masked language modeling objective, allowing the model to predict clean audio tokens of DAC codec~\cite{kumar2024high} conditioned on the degraded mel-spectrogram. We use the pre-trained checkpoint in the original work, which comprises $\sim$360M parameters and is trained on a large-scale dataset comprising 20k hours of speech and singing data.

\noindent\textbf{Autoregressive (AR) Model.}
Our model for AR paradigm, which we term \textbf{\arse}, follows a two-stage pipeline inspired by the architecture introduced in MaskGCT~\cite{wang2024maskgct,amphion}. The first stage employs a 0.5B parameter autoregressive model, initialized from the weights of Qwen2.5-0.5B-Instruct\footnote{\url{https://huggingface.co/Qwen/Qwen2.5-0.5B-Instruct}}. This model takes the noisy audio's dense features from a pre-trained w2v-bert-2 model\footnote{\url{https://huggingface.co/facebook/w2v-bert-2.0}} at the 17th-layer as input and is trained to predict the semantic tokens of clean audio extracted from MaskGCT's semantic codec\footnote{\url{https://huggingface.co/amphion/MaskGCT/tree/main/semantic\_codec}}. We use MaskGCT's second stage\footnote{\url{https://huggingface.co/amphion/MaskGCT/tree/main/s2a\_model}}, which is a non-autoregressive Soundstorm-based model~\cite{borsos2023soundstorm} with approximately 300M parameters. This model converts the predicted semantic tokens into acoustic tokens, which are then synthesized into the final clean audio waveform.

\noindent\textbf{Flow-Matching (FM) Model.}
Our FM model, named \textbf{\flowse}, is designed to learn a continuous mapping from degraded to clean audio representations. It follows the Diffusion Transformer (DiT)~\cite{peebles2023scalable} architecture, which adapts transformer-based models for continuous-time generative tasks. The core of the model consists of 10 LLaMA-2-style decoder layers with 1024 hidden dimensions and 16 attention heads, comprising approximately 220M parameters. The model is also conditioned on dense features of noisy audios from the 17th layer of the pre-trained w2v-bert-2 model and is trained using a flow-matching objective~\cite{lipman2022flow} to predict the clean mel-spectrogram (at a 24kHz sampling rate). For the final waveform synthesis, we employ a pre-trained Vocos vocoder~\cite{siuzdak2023vocos}, specifically leveraging the open-source Vevo1.5 implementation\footnote{\url{https://huggingface.co/amphion/Vevo1.5/tree/main/acoustic\_modeling/Vocoder}}~\cite{zhang2025vevo,amphion_v0.2}.

\subsection{Preference Dataset Construction}

To construct our \tool dataset, we first generated a diverse pool of candidate outputs from the three base models. This diversity is crucial for creating meaningful preference pairs and ensuring the robustness of the subsequent DPO training, we detail the specific stochastic sampling configurations used for each model paradigm below:

\begin{itemize}[leftmargin=*, noitemsep]
    \item \textbf{MGM:} We varied the number of decoding steps by randomly sampling from the range [10, 40]. We also employed nucleus sampling with a fractional top-k value of 0.9 and a temperature of 1.0 to encourage diversity.
    \item \textbf{AR:} For the autoregressive model, we used a combination of sampling techniques: top-k sampling with k=25, nucleus sampling (top-p) with p=0.8, and a temperature of 1.0.
    \item \textbf{FM:} We introduced stochasticity by varying the number of sampling steps (randomly chosen from [10, 40]) and by using different Classifier-Free Guidance (CFG) scales, which were randomly sampled from the range [1.0, 3.0].
\end{itemize}

\subsection{Training Details}

\noindent\textbf{Pre-Training Data:} For \arse and \flowse, we sampled 7k hours of clean speech from the Emilia dataset~\cite{he2024emilia}. And we use the same noise \& rir introduced in \anyenhance~\cite{zhang2025anyenhance} which contains $\sim$1.2k hours noise and 62,668 rir files, to generate the noisy and reverberant audios.

\noindent\textbf{Pre-Training Procedure:} We both train \arse and \flowse using the Adam~\cite{kingma2014adam} optimizer, with a learning rate of 5e-4 and 1e-4 respectively and 32K warmup steps, following the inverse square root learning schedule. The training is performed on 4 GPUs, with each GPU processing 5000 frames of audio per step, which is equivalent to 100 seconds of audio, for 200k steps. The training takes approximately 2.5 days for both two models.

\noindent\textbf{DPO-Training Procedure:} We also use Adam optimizer for DPO training, with different hyperparameters for different paradigms, all models are trained on 2 GPUs:

\begin{itemize}[leftmargin=*, noitemsep]
    \item \textbf{\anyenhance:} We use a learning rate of 1e-6 with a warmup of 4k steps and a batch size of 4 pairs of samples (i.e., one positive and one negative sample) per GPU, and train for 60k steps. The $\beta$ in DPO loss is set to 10.
    \item \textbf{\arse:} We use a learning rate of 1e-5 with a warmup of 4k steps and a batch size of 4 pairs of samples per GPU, and train for 40k steps. The $\beta$ in DPO loss is set to 0.1.
    \item \textbf{\flowse:} We use a learning rate of 1e-7 with a warmup of 4k steps and a batch size of 12 pairs of samples per GPU, and train for 100k steps. The $\beta$ in DPO loss is set to 1000.
\end{itemize}

\section{Evaluation Details}
\label{app:evaluation_details}

\subsection{Baseline Models}
Besides comparison with the three models before and after DPO alignment, we also compare with the following baselines:

\begin{itemize}
    \item \textbf{Voicefixer}~\cite{liu2022voicefixer}: A discriminative general speech restoration model, consists of a ResUNet-based analysis stage which predicts a ratio mask over the degraded mel-spectrogram, and a synthesis stage which converts the masked mel-spectrogram into the final waveform using TFGAN-based vocoder. We use its pre-trained checkpoint\footnote{\url{https://github.com/haoheliu/voicefixer}}.
    \item \textbf{MaskSR}~\cite{li2024masksr}: A masked generative model for speech restoration tasks, which predicts acoustic tokens from DAC codec~\cite{kumar2024high} conditioned on the noisy mel-spectrogram. We report the results in the original paper.
    \item \textbf{GenSE}~\cite{anonymous2025gense}: A two-stage autoregressive generative framework for speech enhancement, which decouples denoising and generation by first mapping noisy speech to clean semantic tokens and then generating clean acoustic tokens. We report the results in the original paper.
    \item \textbf{LLaSE-G1}~\cite{kang2025llase}: A LLaMA-based generative speech enhancement model that preserves acoustic consistency by predicting fine-grained speech tokens from continuous WavLM features, and supports diverse SE tasks like denoise, echo cancellation, etc. We report the results in the original paper.
    \item \textbf{FlowSE}~\cite{wang2025flowse}: A flow-matching-based speech enhancement model that learns a continuous mapping from noisy mel-spectrogram to clean mel-spectrogram, with optional text input conditioning capability. We report the results in the original paper.
\end{itemize}

\subsection{Evaluation Details}

\noindent\textbf{Metrics:} We adopt the following objective and subjective metrics, which are designed to assess various aspects of enhancement quality, including perceptual quality, signal fidelity, and content consistency and timbre preservation:

\begin{itemize}
    \item \textbf{DNSMOS}~\cite{reddy2022dnsmos}: a reference-free metric for 16 kHz audio, providing \textbf{SIG}, \textbf{BAK}, and \textbf{OVRL} scores (1-5) for signal quality, background noise, and overall quality.
    \item \textbf{NISQA}~\cite{mittag2021nisqa}: a reference-free metric for 48 kHz audio that outputs a single overall perceptual quality score (1-5).
    \item \textbf{SpeechBERTScore}~\cite{saeki2024speechbertscore}: evaluates semantic similarity between enhanced and reference signals using HuBERT-base\footnote{\url{https://huggingface.co/facebook/hubert-base-ls960}}.
    \item \textbf{Similarity}: measures speaker similarity via cosine distance between WavLM-based embeddings\footnote{\url{https://huggingface.co/microsoft/wavlm-base-plus-sv}}.
\end{itemize}

\begin{figure}[htbp]
    \centering
    \includegraphics[width=0.99\linewidth]{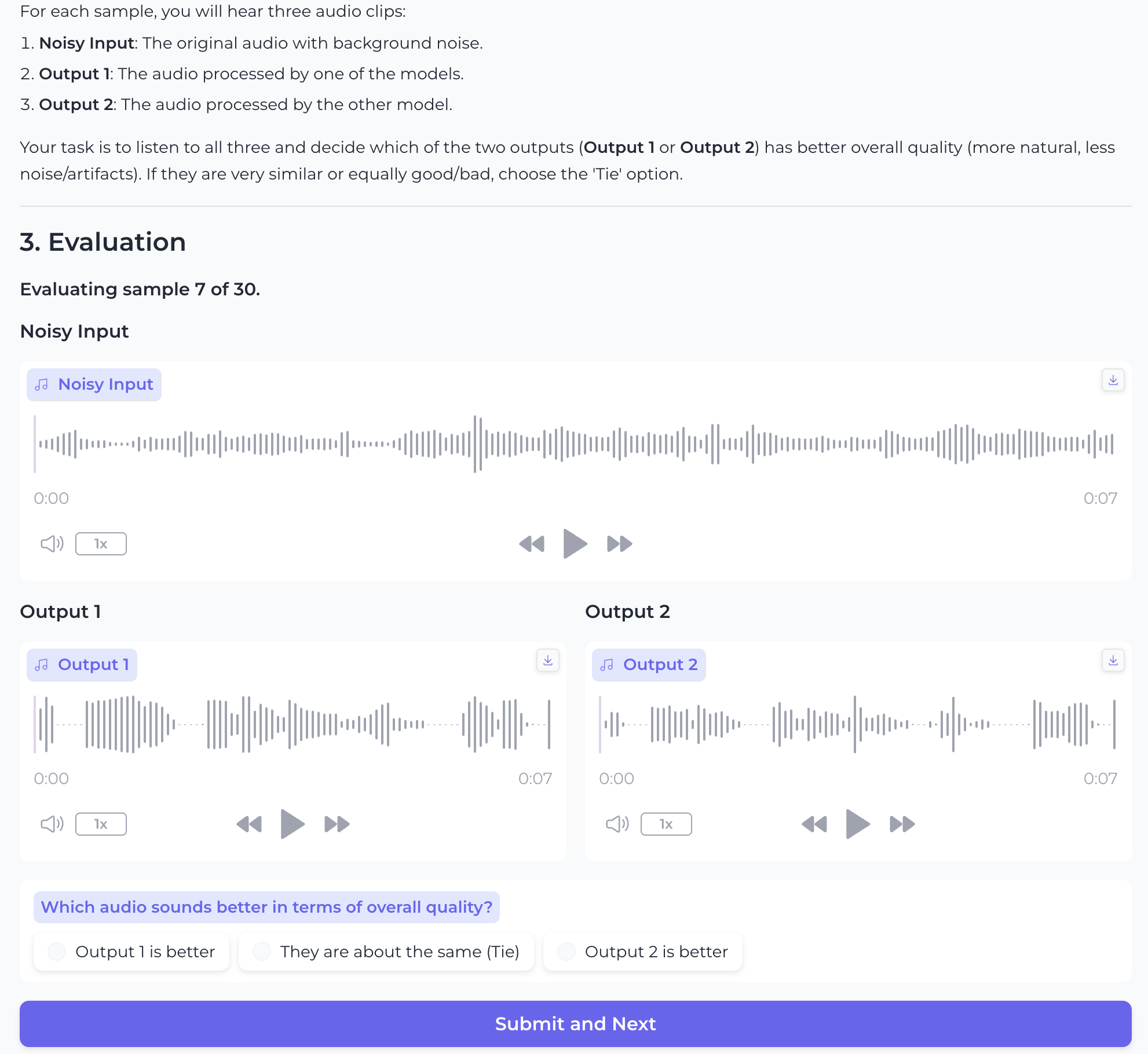}
    \caption{AB Test interface for subjective evaluation.}
    \label{fig:abtest_interface}
\end{figure}

\noindent\textbf{Subjective Evaluation} We conduct subjective evaluation using an AB test, where participants are asked to choose the preferred audio sample between two options. The interface is shown in Figure~\ref{fig:abtest_interface}. Each audio pair was evaluated by 10 raters, the majority of whom are researchers working in speech generation and restoration.

\noindent\textbf{Evaluation Testsets.}
We assess our models on a diverse range of benchmarks covering two primary restoration scenarios: General Speech Restoration (GSR) and Speech Enhancement (SE).
\begin{itemize}
    \item \textbf{General Speech Restoration (GSR):} To evaluate performance on complex, simultaneous distortions (e.g., noise, reverberation, clipping, and bandwidth limitation), we use three test sets: the official Voicefixer-GSR dataset~\cite{liu2022voicefixer}, and two simulated sets introduced in \anyenhance based on clean speech from LibriVox~\cite{kearns2014librivox} and clean singing from CCMusic~\cite{ccmusic}.
    
    \item \textbf{Speech Enhancement (SE):} To evaluate performance specifically on downstream SE tasks, we use the official test sets from the DNS Challenge 2020~\cite{reddy2020interspeech}, including both the with-reverb and no-reverb conditions.
\end{itemize}

\begin{figure*}[h]
    \centering
    \includegraphics[width=\linewidth]{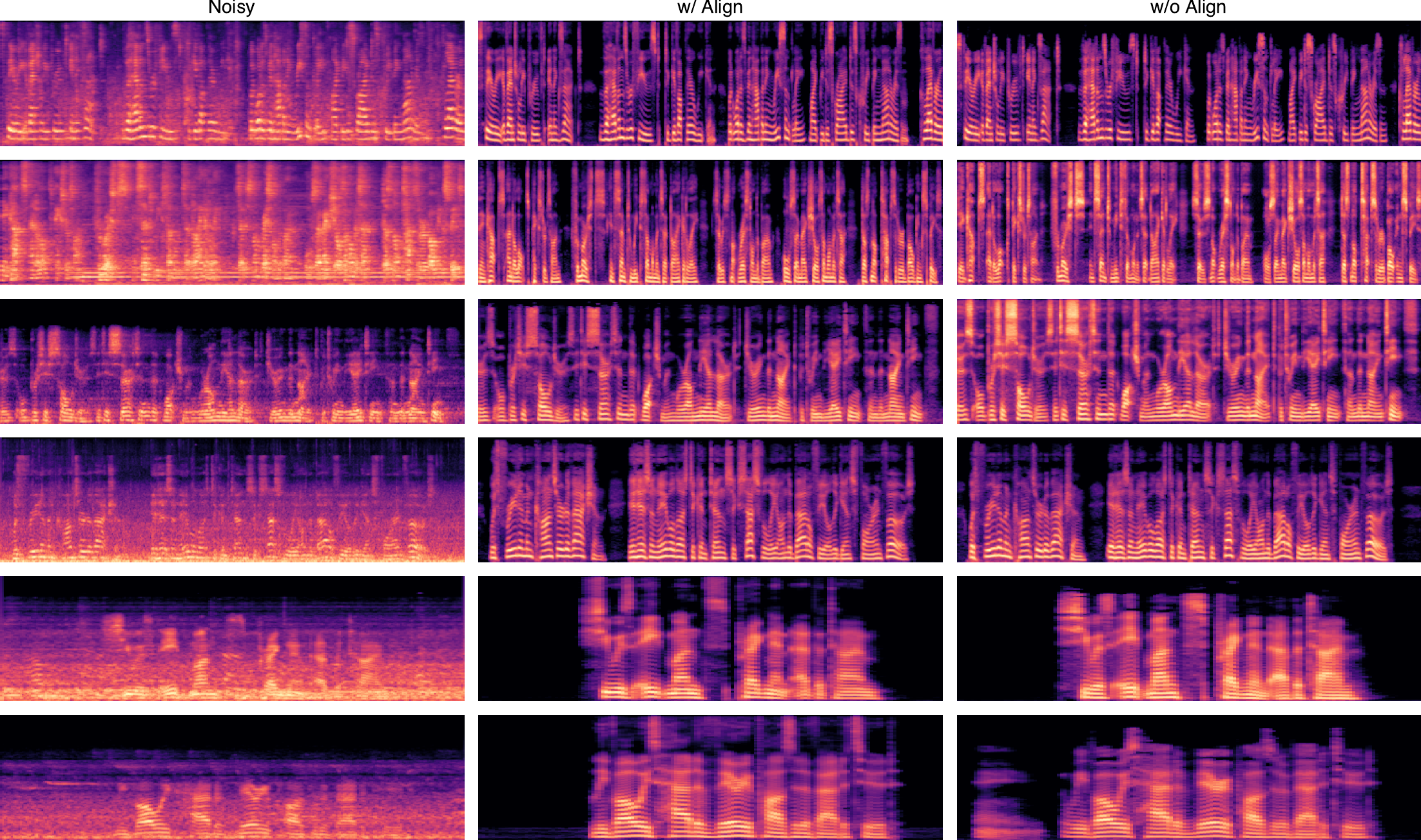}
    \caption{
        Spectrogram comparison for the \textbf{AR+Soundstorm (AR)} model. Each row compares the original noisy audio (left), the restored audio from our DPO-aligned model (center), and the output from the baseline model without alignment (right).
    }
    \label{fig:ar_visualization}
\end{figure*}

\section{More Experimental Results}
\label{app:more_results}

\subsection{Full Results on DNS Challenge Benchmarks}
\label{app:full_results_dns}
Table~\ref{tab:results-se} presents the detailed results on the DNS Challenge benchmarks. Our findings confirm that the DPO-aligned models consistently outperform their original counterparts across both non-reverberant and reverberant conditions. Specifically, we observe significant gains in DNSMOS scores (SIG, BAK, OVRL) and substantial boosts in perceptual quality (NISQA) for all three paradigms. Interestingly, while the alignment strongly favors perceptual and signal quality, there are occasional minor trade-offs in SpeechBERTScore and Speaker Similarity, suggesting the model learns to prioritize clarity and noise reduction, which aligns well with the goals of speech enhancement.

\subsection{AB Test on CCMusic-GSR Testset}
\label{app:abtest_ccmusic}

\begin{figure}[h]
    \centering
    \includegraphics[width=0.9\linewidth]{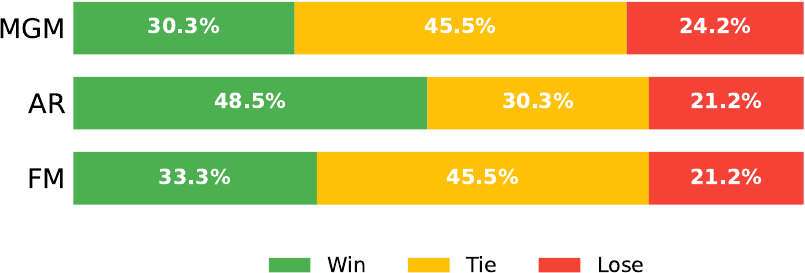}
    \caption{
        Human AB preference test results for DPO-aligned models on CCMusic-GSR testset.
    }
    \label{fig:abtest_ccmusic}
\end{figure}

To further validate the effectiveness of our DPO-aligned models, we also conducted an AB preference test on the CCMusic-GSR testset. The results, shown in Figure~\ref{fig:abtest_ccmusic}, show consistent improvements across all paradigms, with win rates of 30.3\% (MGM), 48.5\% (AR), and 33.3\% (FM), demonstrating the effectiveness of DPO alignment in the singing voice domain.

\begin{figure*}[h]
    \centering
    \includegraphics[width=\linewidth]{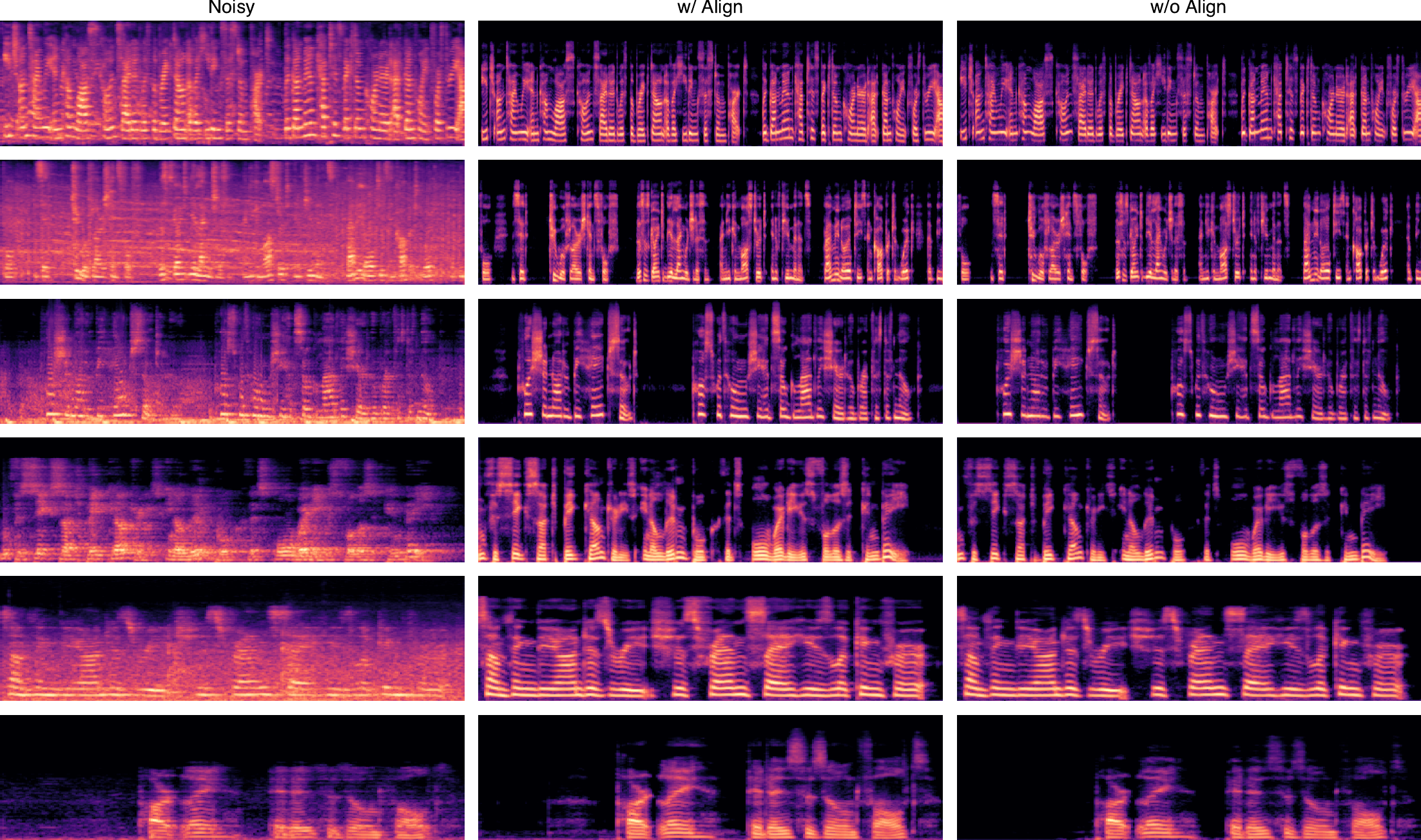}
    \caption{
        Spectrogram comparison for the \textbf{AnyEnhance (MGM)} model. Each row compares the original noisy audio (left), the restored audio from our DPO-aligned model (center), and the output from the baseline model without alignment (right).
    }
    \label{fig:mgm_visualization}
\end{figure*}

\subsection{More Result on Multi-Metric Preference Ablation}
\label{app:multi_metric_ablation}

To fully validate the robustness of our multi-metric alignment strategy, we conducted additional ablation studies on the preference metric for both the FM and MGM models, supplementing the AR model results presented in the main paper. The results, shown in Table~\ref{tab:metric_ablation_fm} and Table~\ref{tab:metric_ablation_mgm}, demonstrate a consistent pattern across all generative paradigms.

\begin{table}[h]
    \centering
    \small
    \setlength{\tabcolsep}{1.2mm}
    \begin{tabular}{lccccccc}
        \toprule
        \textbf{Criterion} & \textbf{SIG} & \textbf{BAK} & \textbf{OVRL} & \textbf{NISQA} & \textbf{SBERT} & \textbf{SIM} \\
        \midrule
        - & 3.398 & 3.969 & 3.104 & 4.010 & 0.812 & 0.918 \\
        Multi-Metric & 3.483 & 4.092 & 3.230 & 4.672 & 0.830 & 0.924 \\
        NISQA & 3.415 & 4.035 & 3.147 & 4.495 & 0.814 & 0.920 \\
        OVRL & 3.464 & 4.010 & 3.179 & 4.206 & 0.815 & 0.918 \\
        SBERT & 3.413 & 4.002 & 3.129 & 4.209 & 0.826 & 0.919 \\
        SIM & 3.394 & 3.951 & 3.091 & 4.095 & 0.810 & 0.921 \\
        \bottomrule
    \end{tabular}
    \caption{Ablation on preference metric for the FM model on Voicefixer-GSR testset.}
    \label{tab:metric_ablation_fm}
\end{table}

\begin{table}[h]
    \centering
    \small
    \setlength{\tabcolsep}{1.2mm}
    \begin{tabular}{lcccccc}
        \toprule
        \textbf{Criterion} & \textbf{SIG} & \textbf{BAK} & \textbf{OVRL} & \textbf{NISQA} & \textbf{SBERT} & \textbf{SIM} \\
        \midrule
        - & 3.406 & 4.073 & 3.136 & 4.308 & 0.829 & 0.924 \\
        Multi-Metric & 3.532 & 4.091 & 3.267 & 4.639 & 0.834 & 0.935 \\
        NISQA & 3.429 & 4.096 & 3.177 & 4.715 & 0.811 & 0.930 \\
        OVRL & 3.536 & 4.088 & 3.274 & 4.459 & 0.820 & 0.928 \\
        SBERT & 3.407 & 4.031 & 3.121 & 4.364 & 0.861 & 0.929 \\
        SIM & 3.372 & 4.060 & 3.107 & 4.381 & 0.841 & 0.936 \\
        \bottomrule
    \end{tabular}
    \caption{Ablation on preference metric for the MGM model on Voicefixer-GSR testset.}
    \label{tab:metric_ablation_mgm}
\end{table}

As with the AR model, aligning with a single metric (e.g., NISQA or OVRL) on the FM and MGM models improves the targeted metric but often leads to stagnation or even degradation in other, non-targeted quality dimensions. For example, in Table~\ref{tab:metric_ablation_mgm}, while the SIM-aligned MGM model improves speaker similarity (SIM), its signal quality scores (SIG and OVRL) fall below the original baseline.  In contrast, the \textbf{Multi-Metric} alignment strategy consistently yields balanced and comprehensive improvements across nearly all evaluation metrics for both models. This reinforces our core hypothesis that a holistic preference signal, derived from a unanimous agreement across a diverse suite of metrics, is crucial for mitigating reward hacking and achieving genuinely superior restoration quality, regardless of the underlying generative architecture.

\subsection{Application: Empowering Discriminative Models via Pseudo-Labeling}
\label{app:application_pseudo_labeling}

To showcase the practical utility of our aligned models, we explore their potential as ``data annotators'' in a challenging, real-world scenario: singing voice restoration, where paired clean/noisy data is often scarce.

\noindent\textbf{Dataset.} We use a proprietary dataset of 183 song recordings. The dataset features 10 different singers and was captured with a variety of recording devices (including smartphones and condenser microphones) across diverse acoustic environments such as piano rooms, rehearsal halls, concert halls, classrooms, bedrooms, and even reverberant spaces like hallways, stairwells, and basements. We partitioned this dataset into a training set of 170 songs (6.97 hours, 3,121 audio clips) and a test set of 13 songs (0.80 hours, 416 audio clips).

\noindent\textbf{Pseudo-Label Generation and Fine-tuning.} We first prepared a specialized \anyenhance model, which was pre-trained with additional singing voice data via the SingNet pipeline~\cite{gu2025singnet} and then aligned on singing voice data with our multi-metric preference alignment strategy. This expert generative model was then used to process the noisy recordings from our training set, generating high-quality pseudo-clean labels for each segment.

\begin{figure*}[!htbp]
    \centering
    \includegraphics[width=\linewidth]{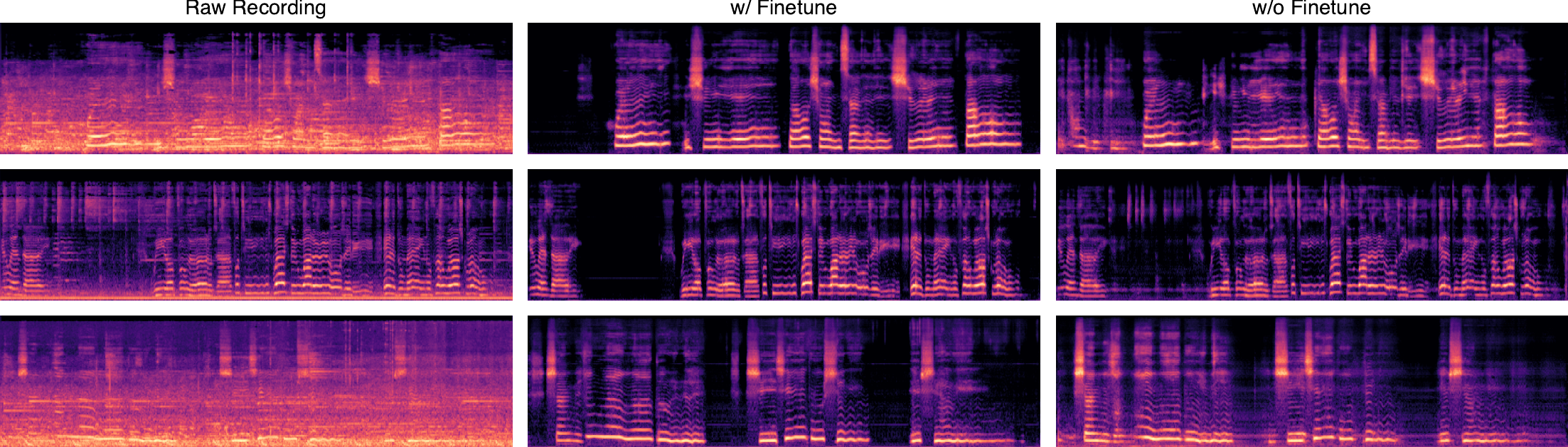}
    \caption{
        Visualization of the singing voice restoration application, demonstrating our aligned model as a ``data annotator''. Each row compares the raw recording (left), the output from the Voicefixer model fine-tuned on our pseudo-labels (center), and the output from the original, un-tuned Voicefixer model (right). The fine-tuned version shows remarkable improvements in clarity and artifact removal.
    }
    \label{fig:singing_visualization}
\end{figure*}

Subsequently, we fine-tuned the official Voicefixer checkpoint on these pseudo-labeled pairs. The fine-tuning was conducted for 5k steps using the Adam optimizer with a learning rate of 1e-5 and a batch size of 24. As shown in \todo{Section~\ref{sec:application}} in the main paper, this approach significantly boosts the performance of the original Voicefixer model. Visualizations of the enhancement quality can be seen in Figure~\ref{fig:singing_visualization}.

\begin{figure*}[h]
    \centering
    \includegraphics[width=\linewidth]{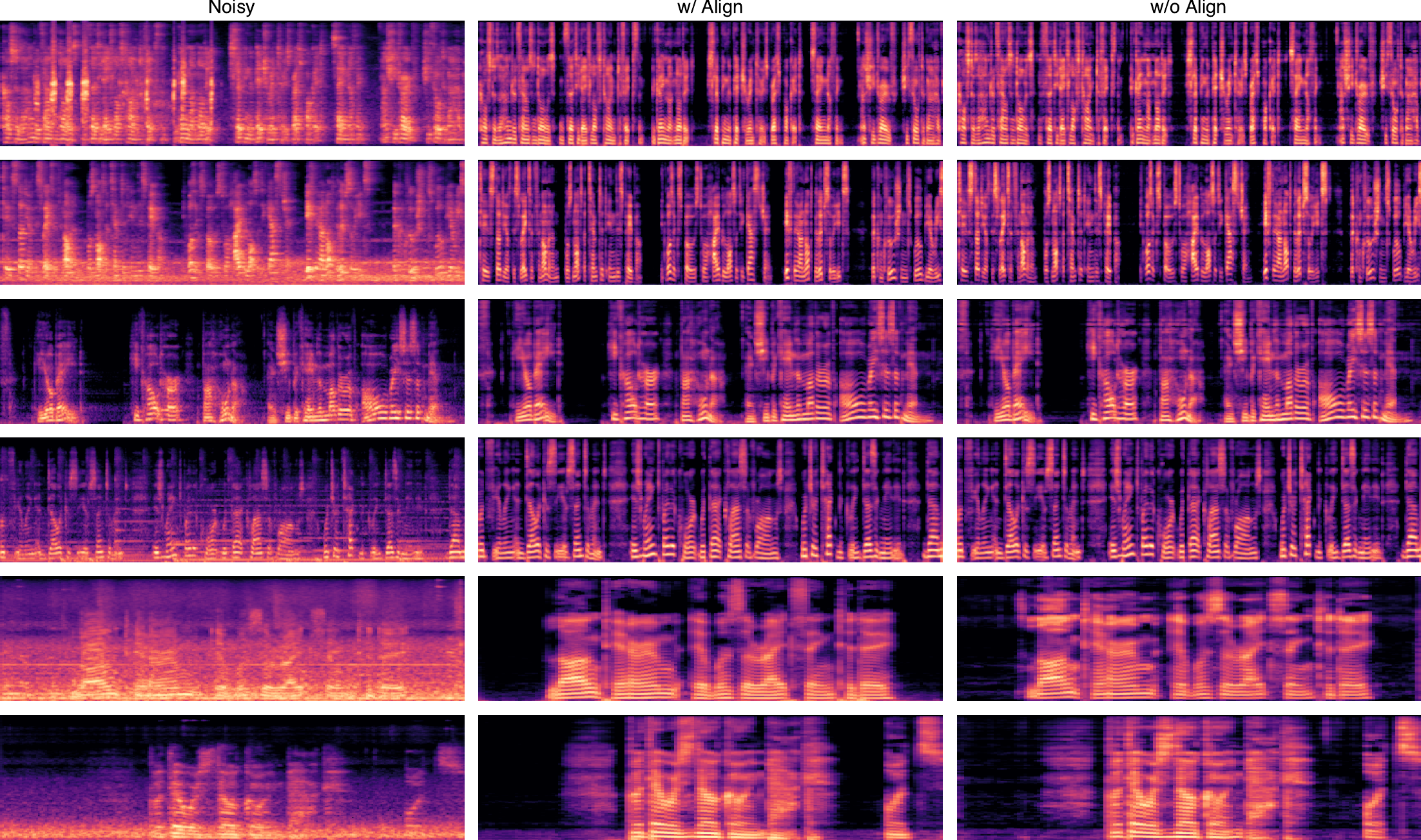}
    \caption{
        Spectrogram comparison for the \textbf{Flow-SR (FM)} model.  Each row compares the original noisy audio (left), the restored audio from our DPO-aligned model (center), and the output from the baseline model without alignment (right).
    }
    \label{fig:fm_visualization}
\end{figure*}

\subsection{Audio Samples Visualization}
\label{sec:appendix_visualization}

To provide a qualitative and visual demonstration of our alignment strategy's impact, we present spectrograms of several audio samples in Figures~\ref{fig:ar_visualization}, \ref{fig:mgm_visualization}, and~\ref{fig:fm_visualization}. 
Each figure corresponds to one of the three generative paradigms. 
Within each figure, every row showcases a direct comparison between the original noisy input (left column), the output from our DPO-aligned model (center column), and the output from the baseline model without alignment (right column).

As can be observed across all examples, the aligned models (\texttt{w/ Align}) consistently exhibit superior performance in suppressing background noise, which appears as a "haze" in the spectrograms of the noisy and baseline outputs. 
Furthermore, our aligned models produce much clearer and more defined harmonic structures (the bright horizontal lines), indicating a higher fidelity restoration of the speech content. This visual evidence strongly corroborates the quantitative improvements reported in the main paper.

Finally, Figure~\ref{fig:singing_visualization} visualizes the results from our application experiment on singing voice restoration. 
This figure demonstrates the practical utility of our aligned models as powerful ``data annotators'' for training discriminative models in data-scarce scenarios, as the fine-tuned Voicefixer model (center column) significantly outperforms the original Voicefixer model (right column) in terms of clarity and artifact removal.

\end{document}